\documentclass[preprint,12pt]{aastex}
\usepackage{natbib}

\slugcomment{\footnotesize {\LaTeX}ed at \number\time\ min., \today}
\shorttitle{SCUBA Mapping of c2d Cores}
\shortauthors{Young et al.}

\begin{document}

\title {SCUBA Mapping of Spitzer c2d Small Clouds and Cores} 

\author{Chadwick H. Young}
 \affil{Department of Physical Sciences, Nicholls State University,
Thibodaux, Louisiana 70301}

\author{Tyler L. Bourke}
\affil{Harvard-Smithsonian Center for Astrophysics, Cambridge, MA 02138, USA}

\author{Kaisa E. Young}
 \affil{Department of Physical Sciences, Nicholls State University,
Thibodaux, Louisiana 70301}

\author{Neal J. Evans II}
\affil{Department of Astronomy, The University of Texas at Austin,
       Austin, Texas 78712--1083}

\author{Jes K. J{\o}rgensen}
\affil{Harvard-Smithsonian Center for Astrophysics, Cambridge, MA 02138, USA}

\author{Yancy L. Shirley}
\affil{Steward Observatory, University of Arizona, Tucson, AZ 85721}

\author{Ewine F. van Dishoeck}
\affil{Leiden Observatory, P.O. Box 9513, 2300 RA Leiden, Netherlands}

\author{Michiel Hogerheijde}
\affil{Leiden Observatory, P.O. Box 9513, 2300 RA Leiden, Netherlands}

\begin{abstract}
We present submillimeter observations of dark clouds that are part of
the Spitzer Legacy Program, From Molecular Cores to Planet-Forming
Disks (c2d).  We used the Submillimetre Common User's Bolometer Array
to map the regions observed by Spitzer by the c2d program to create a
census of dense molecular cores including data from the infrared to
the submillimeter.  In this paper, we present the basic data from
these observations: maps, fluxes, and source attributes.  We also show
data for an object just outside the Perseus cloud that was
serendipitously observed in our program. We propose that this object
is a newly discovered, evolved protostar.  Finally, these data are
publicly available at http://peggysue.as.utexas.edu/SIRTF/DATA/.

\end{abstract}


\section{Introduction}

In recent years, significant progress has been made toward a theory
for the formation of low mass stars, which are the likely sites for
the formation of planetary systems.  Maps of submillimeter dust
continuum emission are excellent tracers of the column density and
mass. Dust continuum emission has been used to map both large clouds 
(e.g. \citealt{mot98}; \citealt{joh04}; \citealt{enoch06}) and smaller
cores (e.g., \citealt{shirley00}; \citealt{visser02}; 
\citealt{young03}; \citealt{kirk05}).

Studies of small cores provide complementary advantages to less-biased surveys 
of large clouds. By focusing on regions already known to have dense gas,
we can study their properties in more detail. The large scale surveys can
in turn check the biases that are introduced.
This paper presents data on dust continuum emission toward a large
number of small cores.

For a full analysis, data at submillimeter wavelengths should be combined
with data through the infrared.
The Spitzer Legacy program, From Molecular Cores to Planet-Forming
Disks (c2d), described by \citet{evans03} 
has obtained extremely sensitive observations from 3.6 to
70 \micron\ toward 105 small cores \citep{huard06}. In
order to obtain the full benefit of this database, it is crucial to
add data at longer wavelengths where the dust emission is
optically thin.

We have mapped 38 small cores, which were initially included in the
c2d program, using the Submillimetre Common-User Bolometer Array
(SCUBA) at both 850 and 450 \micron.  Because of time constraints, 15
of these cores were dropped from the c2d program, though some were
covered by the large cloud map of Perseus (\citealt{joergensen06},
\citealt{rebull06}).  Together with other complementary data at
optical and near-infrared wavelengths, our observations provide a
comprehensive catalog of multi-wavelength data for a sample of small
molecular cores.  The goal of this paper is to present the SCUBA data
in a uniform way, consistent with other papers presenting
submillimeter continuum data on the c2d cores and clouds. We leave
detailed analysis to later papers that combine these data with the c2d
data.

\section{Observations and Data Reduction}

\subsection{Sample}

The sample was drawn from those cores originally included in the c2d
observation plan.  In turn, those cores
had been drawn from catalogs of nearby cores that
have molecular line maps \citep{lee99,jijina99,lee01}.  The gas in
these cores is well-studied, mostly by mapping observations of the 1.3
cm line of NH$_3$ \citep{benson89} and the 3 mm line of N$_2$H$^+$
\citep{caselli02}. These cores are all relatively nearby, within 450
pc, so their maps have good spatial resolution.  A few cores have
been observed in the submillimeter before, but not over a region as large as 
that mapped by the Spitzer Space Telescope.  In these observations, we have
matched the SCUBA maps to cover the same area as the Spitzer maps; in
most cases, these maps are $5\arcmin\times5\arcmin$, but they are
sometimes larger.

Table~\ref{table-sources} lists the central right ascension and
declination for each region observed with SCUBA.  This table also has
the core's associated cloud, if relevant, and the distance, using the
standard distances adopted for the c2d project \citep{huard06}.  We
note whether submillimeter emission was detected in our observations
(Y or N).  For those with a submillimeter detection, we list whether
the submillimeter core has an IRAS detection (within 1\arcmin\ of the
submillimeter peak); these IRAS sources were chosen based on the
criteria given by \citet{lee99}.  We also list whether there was an
IRAS source in those maps with no detected submillimeter emission.
For maps with more than one clearly separated source, we give multiple
entries in the IRAS and SCUBA columns (e.g., YNY).  Next, we indicate
whether the core was observed in the cores part of the c2d program (Y)
or covered by the Perseus cloud map (P); the names for these objects
are listed in boldface type.  Table~\ref{table-sources} includes the
3-$\sigma$ limit for each map at both 850 and 450 \micron, followed by
the atmospheric opacity during the observations, again at both
wavelengths.  Because the goal of this project was to cover the same
observed area as in the c2d program, we often observed in weather
conditions that were inadequate for good 450 $\mu$m maps.  This
allowed us to create large maps, but the quality of the 450 $\mu$m
data, as a result, is not sufficient to analyze dust temperature,
emissivity, etc.

We had intended to observe Per9, whose position is in
\citet{caselli02}.  However, we improperly entered the coordinates at
the telescope and mistakenly observed a field that happens to be
centered on 2MASS 0347392+311912.  This field is about 30$\arcsec$
from the edge of the c2d coverage for the Perseus cloud.  We discuss
the nature of this object in Section~\ref{sxn-2mass}.

\subsection{Observations}

We observed these cores with the 15 m James Clerk Maxwell Telescope
(JCMT) beginning in January 2002 and ending February 2004.  We were
granted time during each semester between these dates.  However, the
data collected during the semester 02A and December 2002 to February
2003 were not usable because of technical problems with the JCMT.  
These are the program identification designations for our
observations: M/01B/N13, M/02A/N18, M/02B/N04, M/03A/N08, and
M/03B/N10.

We used the scan mapping technique \citep{holland99}, which is used to
map regions much larger than SCUBA's field of view.  Through this
method, the telescope scans across the field while chopping by 30, 44,
and 68 degrees in the scan direction, which is in both the direction
of right ascension and declination.  The chopping removes sky
variations and DC offsets, and it creates a difference map of the
source, a positive and negative source separated by the chop throw,
which must be restored by software.

During the course of observations, we measured $\tau_{850}$ and
$\tau_{450}$ with frequent skydips, compared these data with
$\tau_{CSO}$, and found good agreement. Finally, we observed several
cores for flux calibration: CRL 618, Mars, Uranus, and IRC10216. We
generally observed at least 2-3 calibrators during an observing shift.

\subsection{Data Reduction}\label{sxn-reduction}
We analyzed and reduced our data with the SCUBA User's Reduction
Facility (SURF) \citep{jenness97}; we also closely followed the
reduction methods of \citet{pierceprice01}.  Our reduction entailed
flatfielding, extinction correction, removal of bad pixels, removal of
baselines, and the coadding and rebinning of the maps.

To correct the data for atmospheric extinction, we measured $\tau_{850}$ and
$\tau_{450}$ from skydips that were completed at the time of each map;
these values for $\tau$ are listed in Table~\ref{table-sources} and
Table~\ref{table-calibrators} for the cores and calibrators, respectively. 

For the removal of bad pixels, we first used SURF's ``despike2'' tool,
which removes spikes from scanmap observations.  This package has 2
criteria for despiking.  The first considers each data point and asks
how disparate it is from the two adjacent data points.  If this
difference is greater than a certain number of standard deviations for
the scan (NSIGMA), the data point is rejected.  We use NSIGMA=4.  The
second criterion convolves the data with a three sample box and uses
NSIGMA in the same way.  In addition, we have used ``sclean'' to
interactively search all of the data for bad pixels.

We use SURF's ``scan\_rlb'' to remove the baselines from the
extinction corrected and cleaned data.  We assume there is no emission
at the end of each scan, which is generally true, and use the linear
method of baseline removal.  In this way, scan\_rlb fits the baseline
to the end of each scan and then applies it to the remaining scan.

Next, we rebin the data, with baselines removed, to create the
dual-beam map. We used pixels appropriate for Nyquist sampling
(3.5$\arcsec$ and 7.5$\arcsec$ for 450 and 850 $\mu$m, respectively)
and do not smooth the maps; smoothing of the maps often causes
artifacts from the data reduction to appear as real emission (greater
than 3$-\sigma$).  At this point, we examine each rebinned, dual-beam
map to search for evidence of bad data, such as stripes or other
artifacts, missed during the despiking and cleaning tasks.

\begin{figure}
\plotone{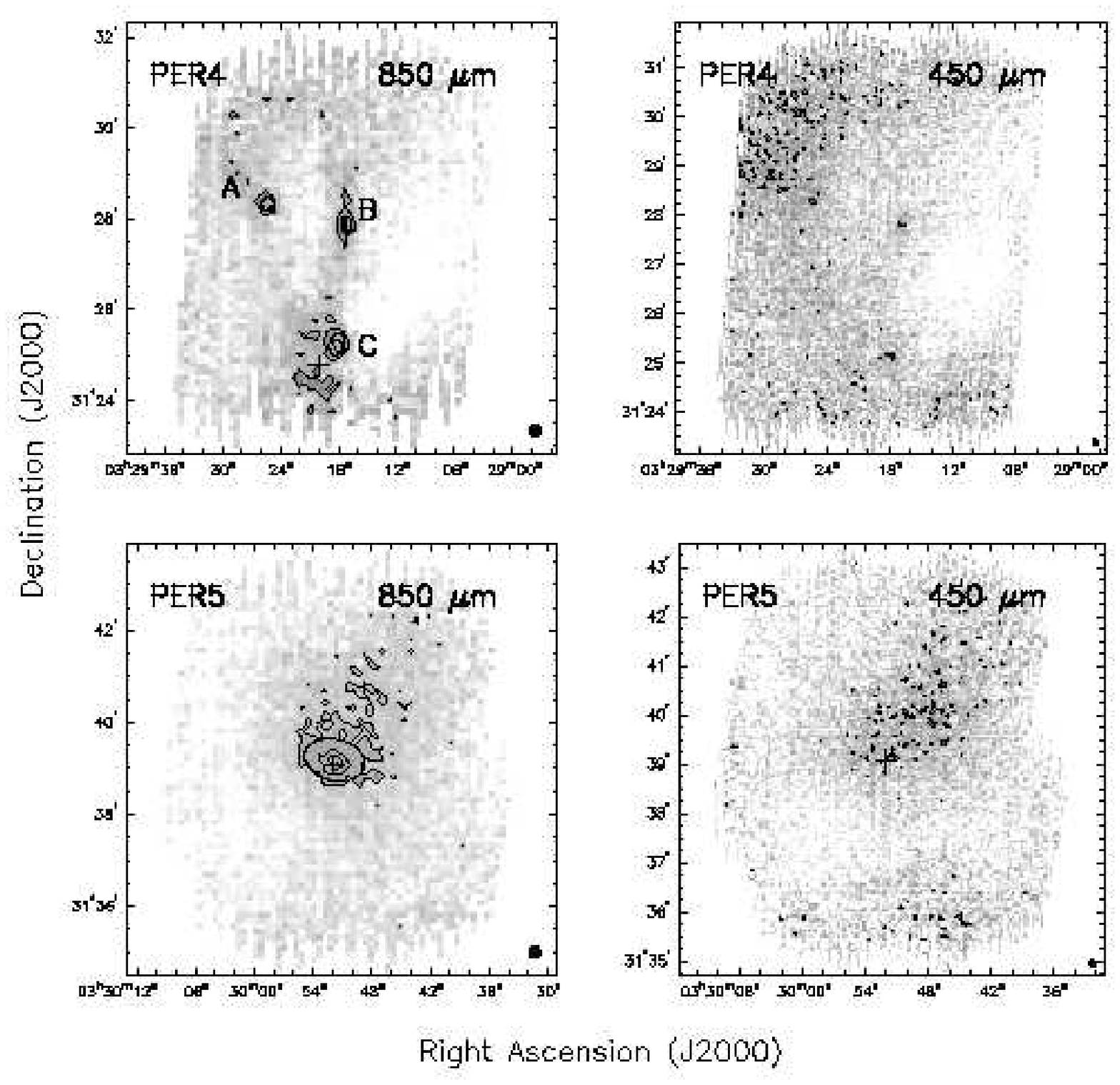}
\figurenum{1} \figcaption{\label{map} We show the 850 and 450 $\mu$m
maps for the  cores in Table~\ref{table-sources}. Contours
begin at 2-$\sigma$ and increase by 2-$\sigma$; greyscale begins at
-2-$\sigma$ and increases to the image maximum. 
Crosses mark the position of IRAS sources in the field. The ellipses 
show the size and shape of the source as determined by the Source Extractor
described in Section~\ref{sxn-reduction}.}
\end{figure}

The rebinned maps are the telescope's dual-beam function convolved
with the sky.  We use ``remdbm'' to remove the dual-beam signature
from the map. This package uses the Emerson2
\citep{emerson95,emerson79} technique to deconvolve the data and
create the single-beam map of the source.  We use no high-frequency
filtering. We assess the RMS noise of the map with Starlink's
``stats'' package and create the maps in Figure~\ref{map}.  The
contours begin at 2-$\sigma$ and increase by 2-$\sigma$; greyscale
begins at $-2$-$\sigma$ and increases to the image maximum.  Crosses
mark the position of IRAS sources, which were selected using the
criteria of \citet{lee99}.

\begin{figure}
\figurenum{2} 
\centering
 \vspace*{7.8cm}
   \leavevmode
   \includegraphics{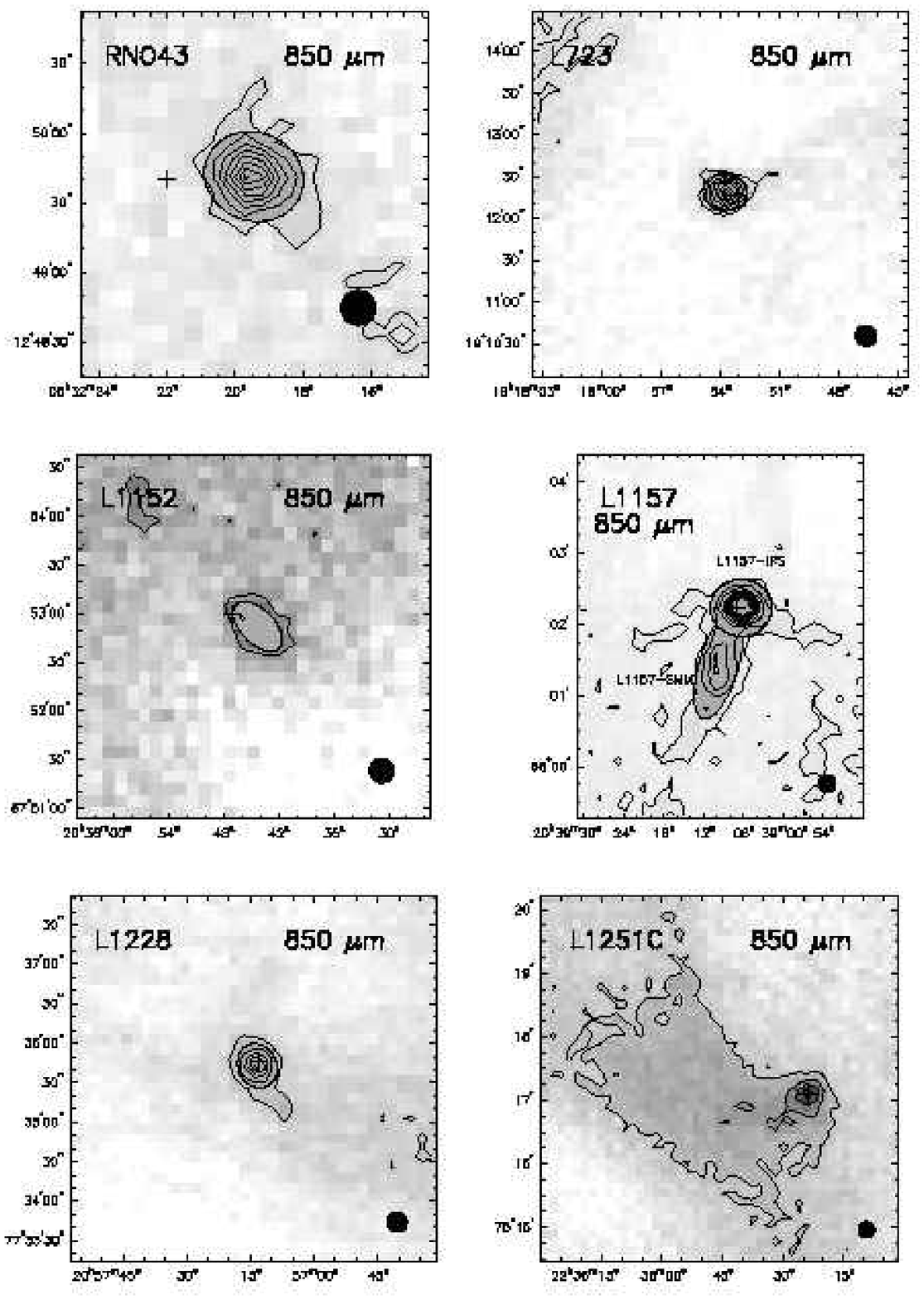}
\vskip 4.25in \figcaption{\label{fig-zoom}These are the central
regions of those maps that cover especially large regions. The
ellipses and crosses are as described for Figure~\ref{map}.}
\end{figure}

We have used Starlink's Extractor, which is based on SExtractor
\citep{bertin03}, to extract source information for each of the maps
in Figure~\ref{map} \citep{chipperfield04}.  This includes the source
positions (barycenter), semi-major and semi-minor axes, and aspect
ratios in Table~\ref{table-properties}.  The aspect ratio is the ratio
of major to minor axes as measured at the 3-$\sigma$ threshold limit
except for L1251C; where we used a threshold limit of 5-$\sigma$.  The
3-$\sigma$ limit, for L1251C, extracted a source that was much larger
than the obviously compact and embedded core.  In Figure~\ref{map}, we
have overlaid, on the maps, an ellipse that matches the extracted
source by Extractor.  For those observations with especially large
fields, maps covering just the central regions are shown in
Figure~\ref{fig-zoom}.

\subsection{Calibration}
In Table~\ref{table-calibrators}, we list flux conversion factors
(FCF) for all calibrators observed during the course of this project.
To calculate the FCF, we divide the flux of the calibrator, which is
assumed to be a point source, by the total data counts (i.e. volts) in
a given aperture.  To determine the FCF for a 20$\arcsec$ aperture,
for example, we divide the flux of the calibrator by the total data
counts for a 20$\arcsec$ aperture, which gives the FCF in units of
Jy/Volt.  Then, we use this FCF to convert the data counts for the
cores into flux units (for a 20$\arcsec$ aperture).  This method of
calibration has been used by a number of authors
\citep[e.g.]{shirley00,young03} and is superior to using beam-sized
apertures because of the significant and variable sidelobes on the
JCMT's beam at 850 and 450 $\mu$m.  These sidelobes, at 850 $\mu$m,
can extend to 50$\arcsec$ even though the beam size is
$\sim$15$\arcsec$ \citep{shirley00}.

In utilization of the FCF, we simply average all values, except those
for Jupiter and IRC 10216, and give this average and standard
deviation in Table~\ref{table-calibrators}.  As suggested on the SCUBA
webpage, we do not use Jupiter and IRC10216 for flux calibration.
(http://www.jach.hawaii.edu/JCMT/continuum/calibration/sens/secondary\_2004.html).
We do not use FCFs when the uncertainty approaches 100\%, as for the
450 $\mu$m FCFs for 80$\arcsec$ and 120$\arcsec$ apertures, but we
list them in Table~\ref{table-calibrators} along with all of the
average calibration factors.

In Table~\ref{table-calibrators}, we also give the FWHM of a Gaussian
fit to the JCMT beam pattern at 850 $\mu$m. We fit a Gaussian to the
observed intensity profile with Starlink's psf routine (setting
gauss=True) and deconvolve the measured beam size for the planet
diameter (Uranus and Mars have semi-diameters of 1\farcs72 and
3\farcs67, respectively).  The average FWHM of the beam, over all
nights, was $15\farcs5 \pm 0\farcs5$.The nominal beam at 450 $\mu$m
is about 8\arcsec, but it can be as large as 11\arcsec
\citep{young03}.  Because the quality of the 450 $\mu$m data is poor,
we do not calculate the 450 $\mu$m beam profile for these observations.

\begin{figure}
\plotone{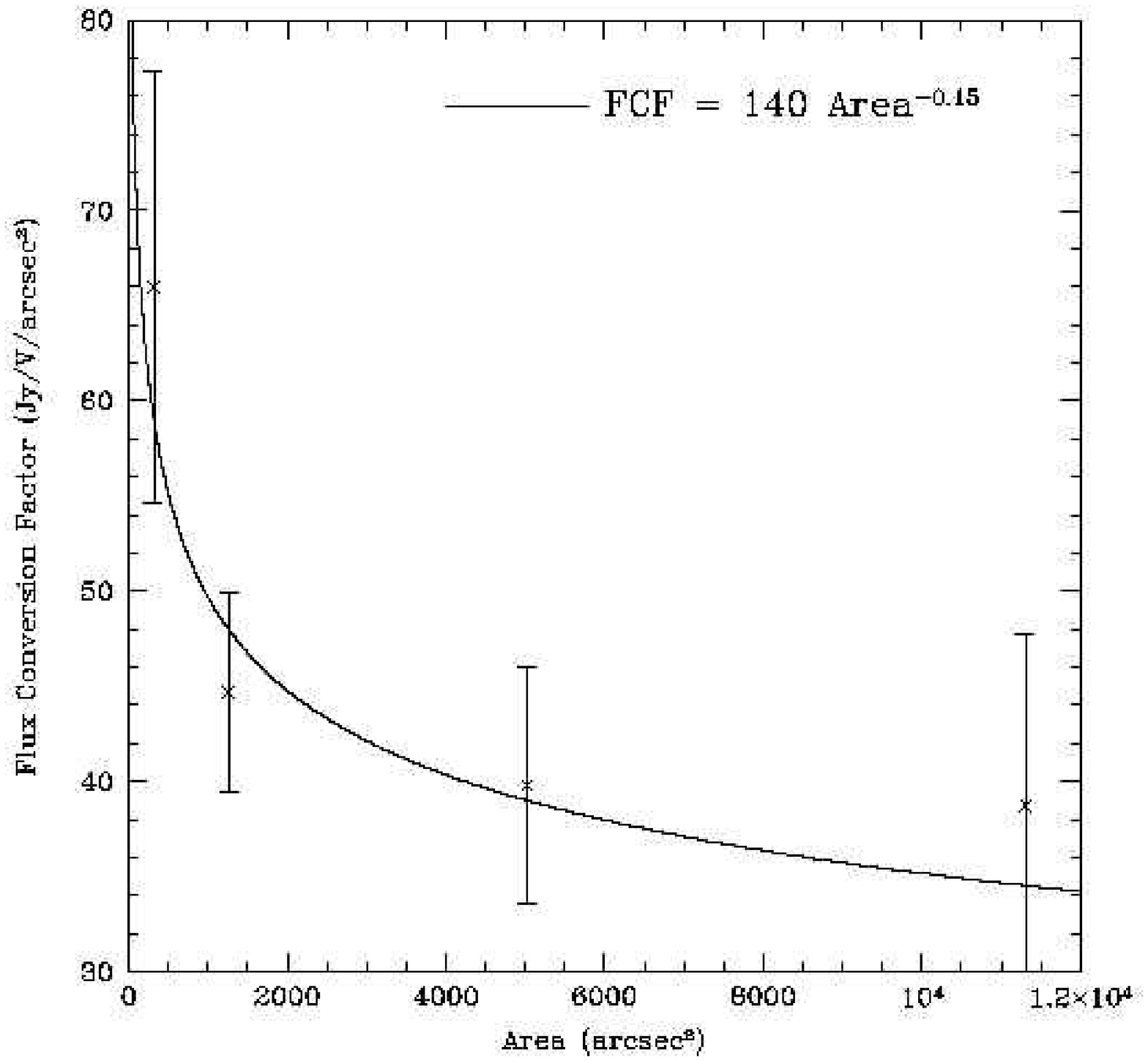} \figurenum{3} \figcaption{\label{fig-calibration} We
have fitted the data for flux conversion factors (shown by the
crosses) with a function in order to calculate the FCF for apertures
with different area.  This function ($FCF=140(Area)^{-0.15}$) is used
in calculating the FCFs used to determine the source fluxes in
Table~\ref{table-properties}.}
\end{figure}

\begin{figure}
\figurenum{4} 
\centering
 \vspace*{7.8cm}
   \leavevmode
   \includegraphics{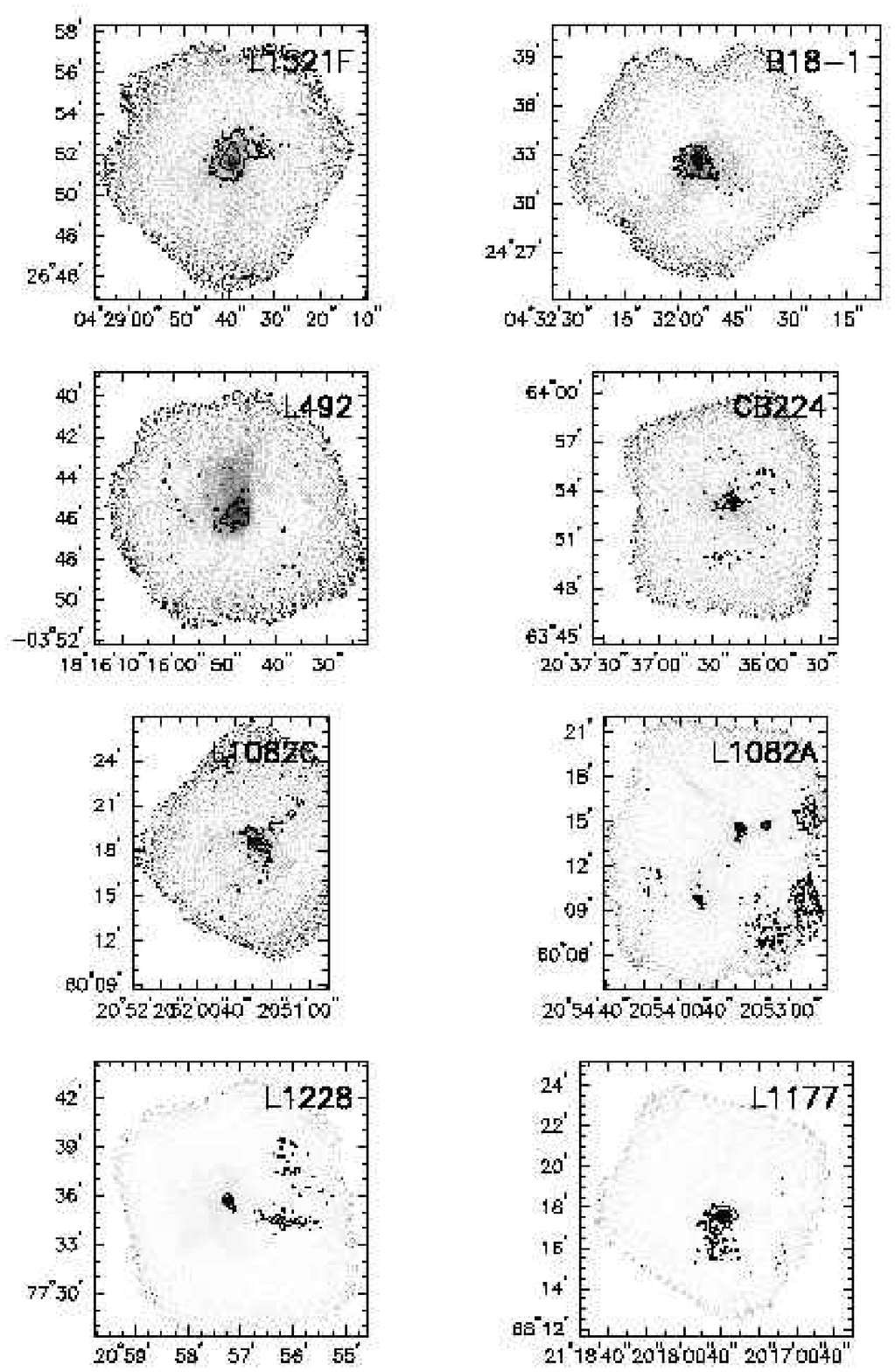}
\vskip 4.25in \figcaption{\label{fig-mambo} MAMBO data for cores are
shown in greyscale \citep{kauffmann06}.  The contours represent 850
$\mu$m observations as shown in Figure~\ref{map}. The SCUBA and MAMBO
data show good agreement on these cores.}
\end{figure}

We have calculated fluxes for each source whose size and shape were
determined by Extractor.  To determine the FCF for these non-circular
apertures, we have fit a function to the FCFs of the circular
apertures.  Because the sidelobes contribute significantly to the
aperture sums, the FCF decreases as the area of the aperture
increases.  In Figure~\ref{fig-calibration}, we show the average FCFs
for the 20\arcsec, 40\arcsec, 80\arcsec, and 120\arcsec\ apertures;
the solid line in this plot is a fit to these data: $FCF=140
Area^{0.15}$, where $Area$ is the area in square arcseconds of the
aperture.  For the non-circular apertures, we calculate the area ($\pi
a b$) and determine the FCF from the above relationship.  Then, we
calculate the flux and ascribe a 30\% uncertainty to this measurement,
which is a liberal estimate of the flux uncertainty at 850 $\mu$m.
The fluxes in the elliptical apertures are in
Table~\ref{table-properties}, and the fluxes in the fixed circular
apertures of different diameters are given in
Table~\ref{table-fluxes}. The apertures are those agreed to for papers
presenting ancillary data to the c2d project.  These ancillary
continuum data are in preparation for publication and include
observations of the c2d cores from Max-Planck-Millimeter-Bolometer
\citep[MAMBO]{kreysa98}, the Submillimeter High Angular Resolution
Camera II \citep[SHARCII]{dowell03}, and the Sest Imaging Bolometer
Array \citep[SIMBA]{nyman01}.  In Figure~\ref{fig-mambo}, we show the
MAMBO data as greyscale with the SCUBA 850 $\mu$m observations
represented by contours, as in Figure~\ref{map}.  There is good
agreement between these sets of data in representing the detected
continuum emission.

\citet{kirk05} observed two of these sources with SCUBA; for L1521F,
they measured the fluxes in a 150$\arcsec$ aperture: 3.23$\pm$0.23 Jy
at 850 $\mu$m and 12.4$\pm$2.4 Jy at 450 $\mu$m.  These are compared
to our measurements of 4.2$\pm$0.7 Jy at 850 $\mu$m (120$\arcsec$
aperture) and 13.3$\pm$7.2 Jy at 450 $\mu$m (40$\arcsec$ aperture).
For L43, \citet{kirk05} measured 5.4$\pm$0.23 and 28.6$\pm$1.8 Jy (at
850 and 450 $\mu$m), which compares to our measured values of
4.8$\pm$1.2 and 15.0$\pm$8.1 Jy.

\section{Results}

In this section, we give the basic results for these observations:
maps, fluxes, and characteristics of the cores.  Further analysis of
these data will follow in later papers and will include Spitzer
observations and other ancillary data (near-infrared, millimeter,
etc.).

Of the 38 cores observed with SCUBA, 13 were not detected at either
850 or 450 $\mu$m.  For these cores and the detected objects, we list
the 3-$\sigma$ noise level in Table~\ref{table-sources}, starting with
source PER7A. These cores do not appear in the other tables.  The maps
for the detected cores are shown in Figure~\ref{map}.  For those
observations with especially large fields, maps covering just the
central regions are shown in Figure~\ref{fig-zoom}. In both of these
figures, crosses represent IRAS sources selected by the criteria of
\citet{lee99}, and the ellipses are the sources chosen by Extractor.

\subsection{Size, Shape and Mass}

In Table~\ref{table-properties}, we give the size and aspect ratio of
each core as measured at the 3-$\sigma$ level with Starlink's
Extractor.  We also list the 850 $\mu$m flux within the aperture
described by this size and aspect ratio; the isothermal mass of each
core is derived from that 850 $\mu$m flux, assuming a temperature of 15
K \citep[see Equation 2]{young03}; the dust opacity used in these
calculations is $\kappa_\nu=0.018$ cm$^{2}$g$^{-1}$ (for OH5 dust;
Ossenkopf \& Henning 1994). The assumed temperature is relevant; a
variation from 10 K to 20 K in isothermal temperature causes an
increase in the mass by a factor of 3.3. The sizes are, with one
exception, much larger than the beam, so deconvolution of the beam is
unnecessary.  The very compact core is not included in the plots.

\begin{figure}
\plotone{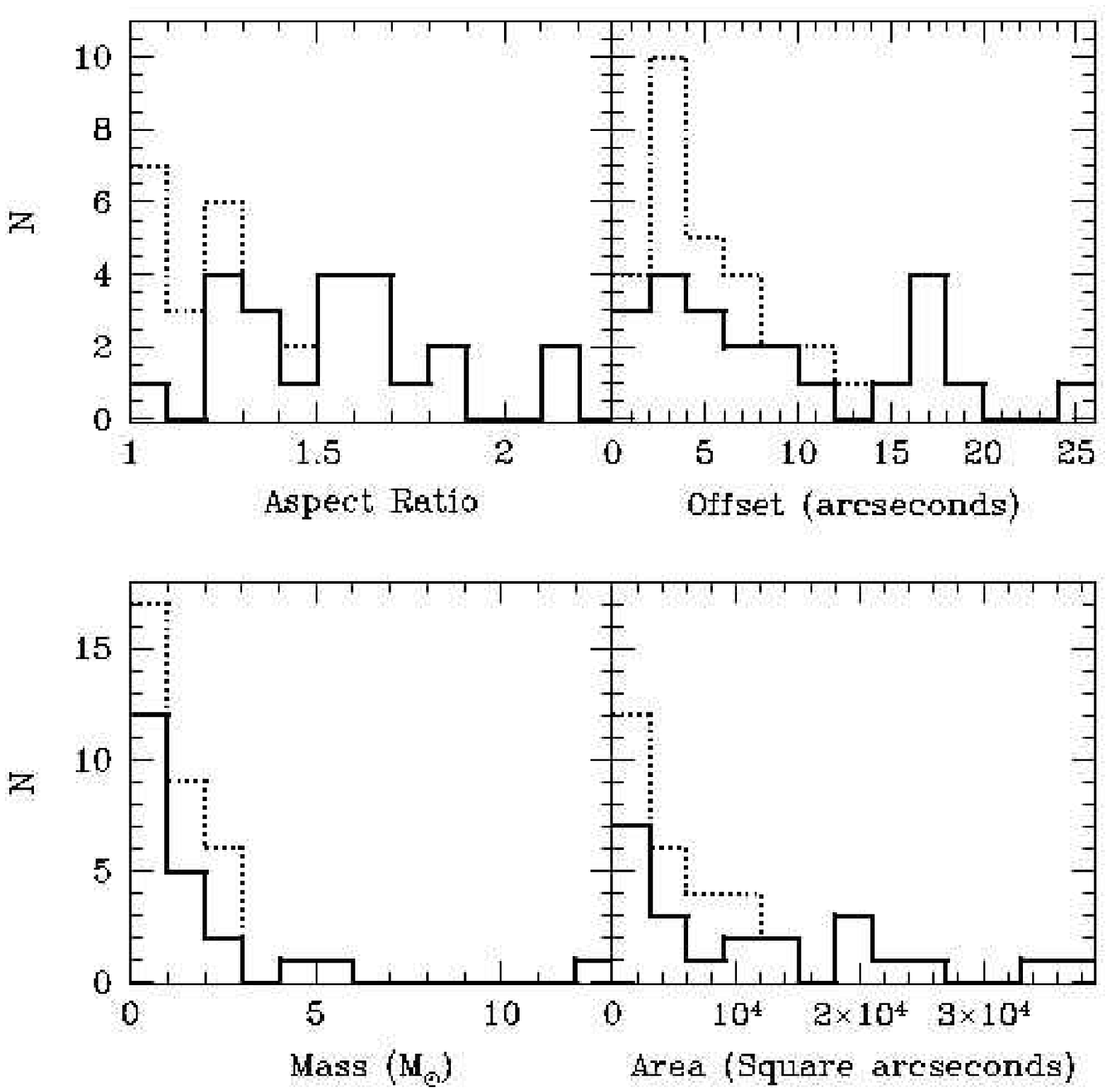} \figurenum{5}
\figcaption{\label{fig-histogram}These histograms show the isothermal
masses, aspect ratios, and areas as reported in
Table~\ref{table-properties}. The solid line represents the data for
cores observed by c2d; the dotted line shows the histogram for all
cores presented in this paper. The area is calculated as $\pi ab$, the
area of an ellipse. The upper right panel shows the difference in
positions of the peak pixel and the barycenter position.}
\end{figure}

In Figure~\ref{fig-histogram}, we plot histograms of the isothermal
masses, aspect ratios, and core areas as reported in
Table~\ref{table-properties}.  The difference in the position of the
peak pixel and the barycenter position (i.e., the center of the fitted
ellipse) is also shown as a histogram in this figure; this is a
measure of axisymmetry.  The core area is equal to $\pi ab$, the area
of an ellipse.  In order to maintain consistency with other c2d
ancillary data papers to come, we show the data for only those cores
observed in the c2d program (i.e. the boldfaced objects in
Table~\ref{table-properties}).  The dotted lines represent the
histogram for the entire sample of cores.

\begin{figure}
\plotone{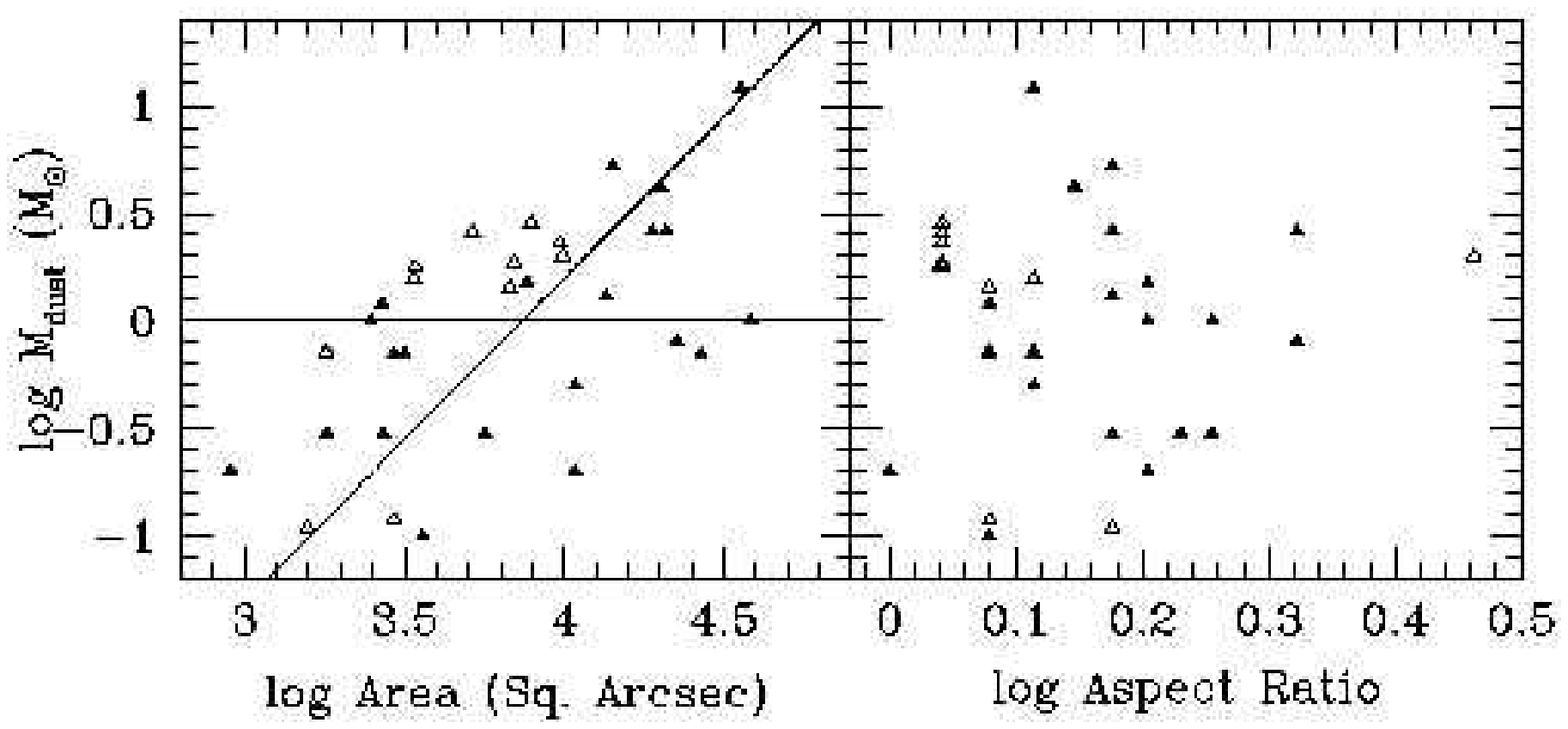} \figurenum{6}
\figcaption{\label{fig-mdust_size} We plot the mass of each core
against the size and aspect ratio as measured at the 3-$\sigma$ level.
The filled triangles are for cores observed by c2d; the open triangles
are for cores not observed by c2d.  To calculate the dust mass, we use
the flux in the elliptical aperture (as shown in Figure~\ref{map}) and
assume an isothermal temperature of 15 K.  The lines in the left panel
represent what is expected for this relationship in cores with
constant column density (flat) and constant density ($M \propto
area^{1.5}$).}
\end{figure}

In Figure~\ref{fig-mdust_size}, we plot the core mass against the size
and aspect ratio of the cores (as measured at the 3-$\sigma$ level);
the data for those cores observed in the c2d program are represented
by filled triangles; the cores not observed by c2d are shown as open
triangles. The lines in the left panel (with area) represent what is
expected for this relationship in cores with constant column density
(flat) and constant density ($M \propto area^{1.5}$).

The dust mass increases with size as one might expect, though the
scatter is large. There are also a number of cores that have low
masses despite their large core size.  Perhaps, these objects are
cold, starless cores.

\subsection{Multiplicity}

Several of the maps contain more than one dense core: IRAM 04191+1522,
L1082A, L1157, L1221, L1251E, L43, and Per4.  Three other objects have
some extended emission, which is either real or an artifact of the
double-beam deconvolution \citep{pierceprice01}: B35A, CB68, and
L1251C. For B35A, at least, the extended emission does appear to be
real because the N$_2$H$^+$ map shows similar structure
\citep{caselli02}. The N$_2$H$^+$ map of L1251C is also similar to the
850 $\mu$m continuum image.  CB68 was mapped by \citet{huard99}, and
their map, while much smaller and more sensitive than ours, does not
appear to show the extended emission in Figure~\ref{map}.

In all but one case, only one object in each of the multiple core
systems was detected by IRAS.  This implies that either these multiple
cores are at different stages of evolution or that some of the cores
are forming objects with lower masses than their companion and hence
are not detected by IRAS.  With the Spitzer Space Telescope, we have
begun to unravel the mystery of these cores.  For example, IRAM
04191+1522 does harbor an infrared source detected with IRAC and MIPS
\citep{dunham06}, but L43-SMM does not appear to harbor an embedded
source down to very low luminosities \citep{huard06}.

\section{2MASS 0347392+311912}\label{sxn-2mass}

We inadvertently observed 2MASS 0347392+311912, having intended to
observe a known core in Perseus instead.  The source lies
outside of the c2d observations for Perseus, so we have no mid- or
far-infrared data.  The 2MASS source, however, did have detectable
submillimeter emission at 850 $\mu$m (0.2 Jy).

\begin{figure}
\plotone{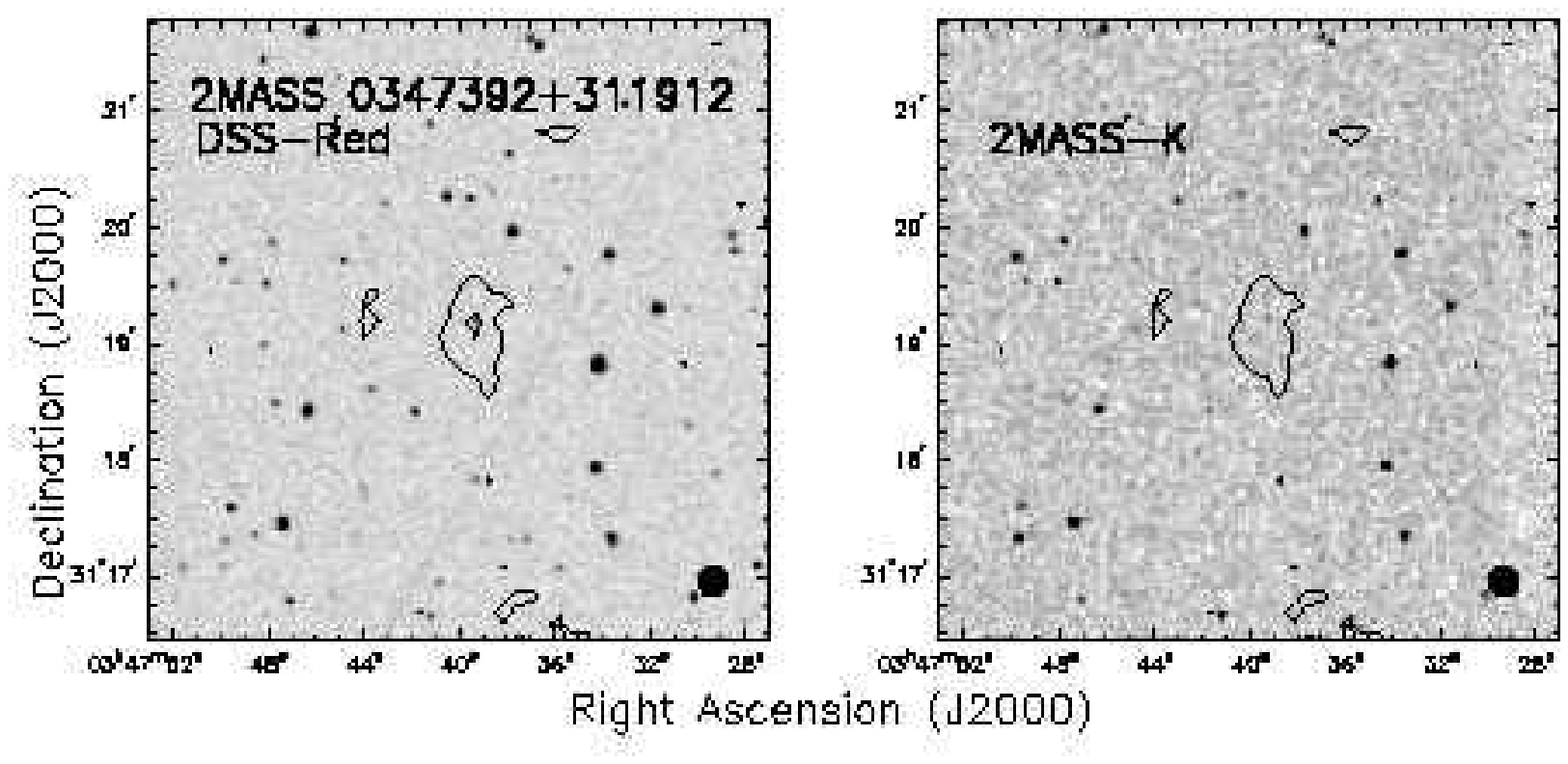} \figurenum{7}
\figcaption{\label{2mass-map}We show data for 2MASS
0347392+311912. The greyscale in the left panel is from the Digitized
Sky Survey, and the right panel shows the 2MASS K-band image.  The
contours are from the SCUBA 850 $\mu$m observations.  Contours begin
at 2-$\sigma$ and increase by 2-$\sigma$; we do not draw the
4-$\sigma$ contour in the right panel because it obscures the 2MASS
source, 2MASS 0347392+311912.}
\end{figure}

\begin{figure}
\plotone{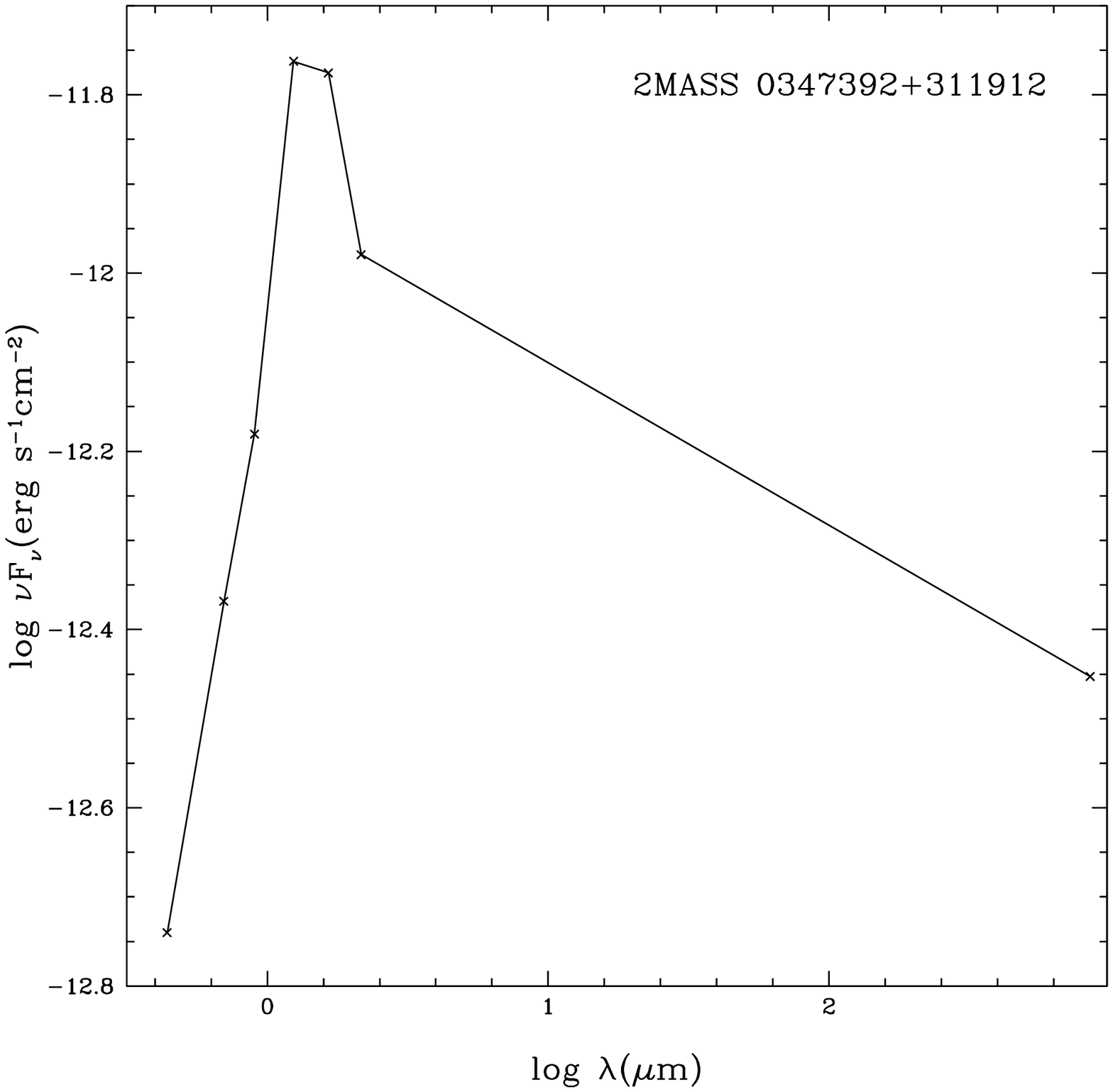}  \figurenum{8}
\figcaption{\label{2mass-sed}We show the SED for 2MASS
0347392+311912. The B, R, and I fluxes are from the USNO-B1.0 Catalog.
J, H, and K data are from 2MASS.  The 850 $\mu$m data is from this
work.}
\end{figure}

In Figure~\ref{2mass-map}, we show the 850 $\mu$m data overlaid on
2MASS K-band and DSS R-band images.  In both the near-infrared and
optical maps, the source appears pointlike and is clearly detected.
In Figure~\ref{2mass-sed}, we plot the SED for this source, including
the near-infrared data from 2MASS and B, R, and I measurements from
the USNO B1.0 catalog.  The SED peaks at 1.65 $\mu$m (H), but and has
substantial submillimeter emission (0.1 Jy at 850 $\mu$m).  This
source may be a previously unknown protostar, which is possibly near
the Perseus molecular cloud.

\section{Summary}

We have presented SCUBA observations of low-mass cores that were
initially proposed to be observed with the Spitzer Space Telescope as
a part of the c2d Legacy Program.  We describe these observations and
the data reduction and analysis.  In addition, we show the maps,
source fluxes, and source sizes.  The core mass has been calculated
from the 850 $\mu$m fluxes, and we analyze how the mass varies with
source size and shape.  We find several small clouds with multiple,
distinct cores, but, for most regions, only one of the cores is
detected by IRAS.  Finally, we present inadvertent observations of an
object near Perseus that may be a newly discovered protostar.  All
data presented here are publicly available as FITS images at
http://peggysue.as.utexas.edu/SIRTF/DATA/.

\section{Acknowledgements}

The authors acknowledge the data analysis facilities provided by the
Starlink Project which is run by CCLRC on behalf of PPARC.  We are
also grateful to the JCMT support scientists Remo Tilanus, Vicky
Barnard, and Douglas Pierce-Price for much needed help in observing
and data reduction.  Support for this work, part of the Spitzer Legacy
Science Program, was provided by NASA through contract 1224608 issued
by the Jet Propulsion Laboratory, California Institute of Technology,
under NASA contract 1407. Astrochemistry in Leiden is supported by a
NWO Spinoza grant and a NOVA grant. This work was also supported by NASA
grants NAG5-10488 and NNG04GG24G.

\begin{figure}
\plotone{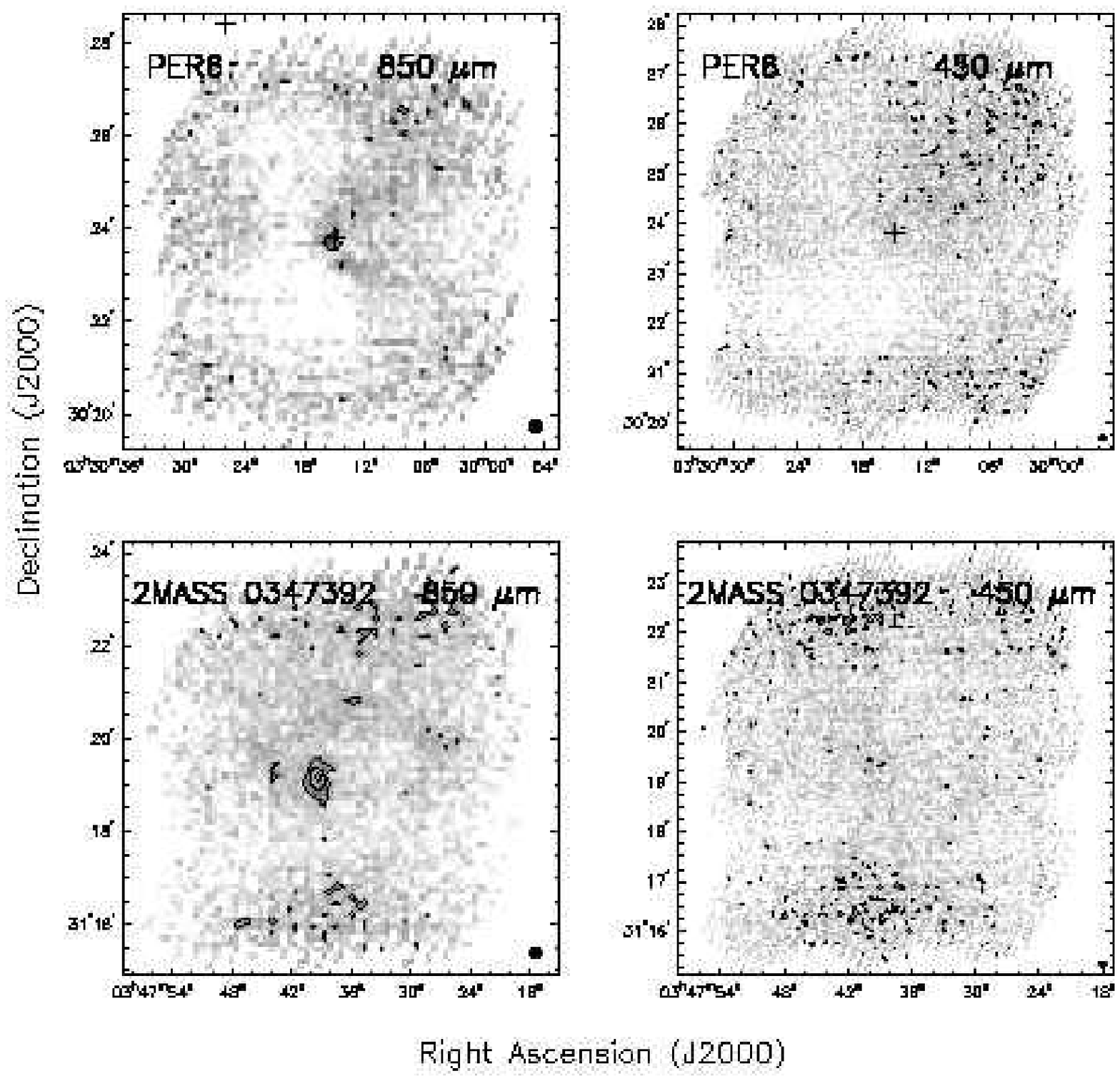} 
\figurenum{1}
\figcaption{continued...}
\end{figure}

\begin{figure}
\plotone{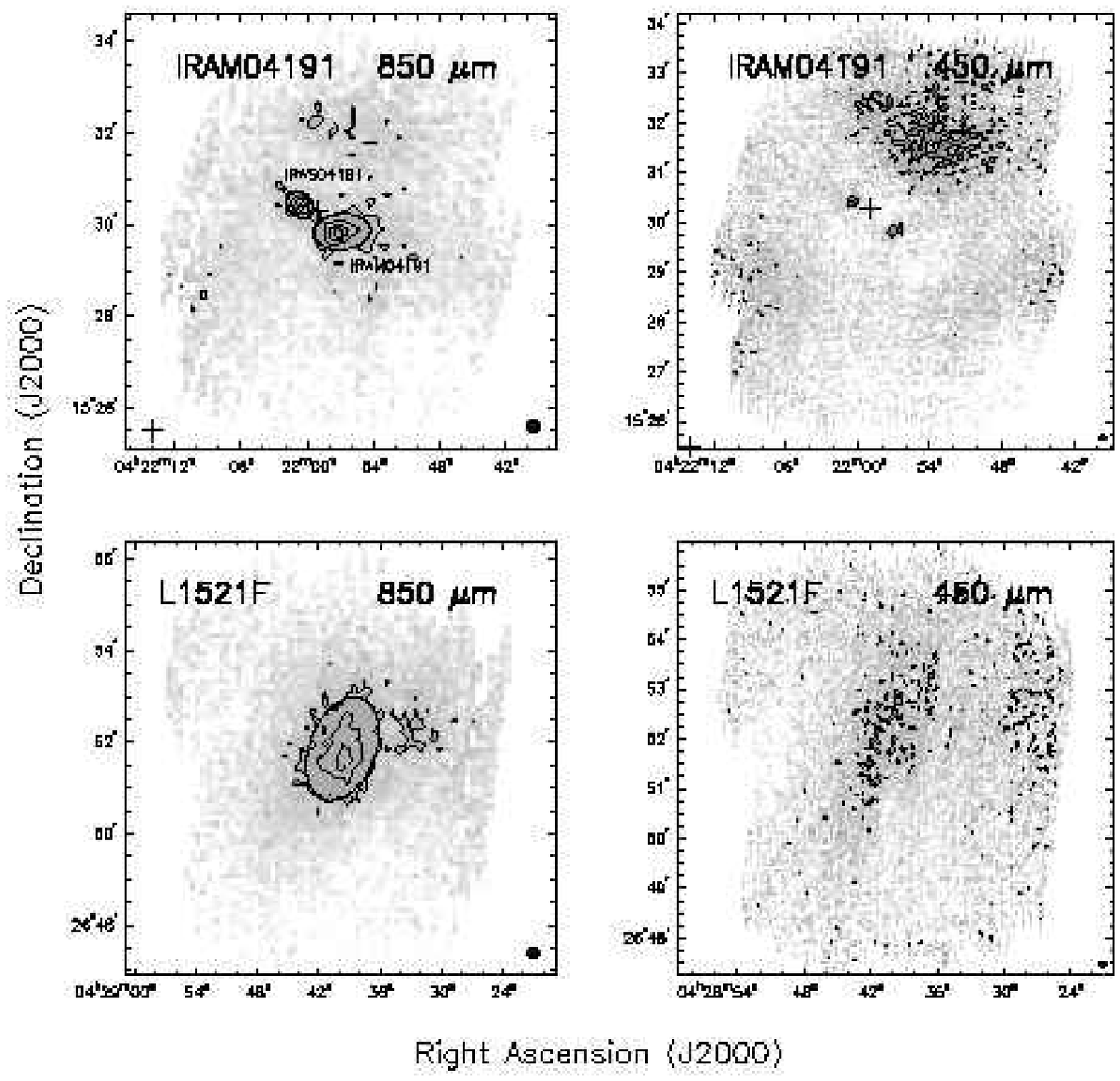} 
\figurenum{1}
\figcaption{continued...}
\end{figure}

\begin{figure}
\plotone{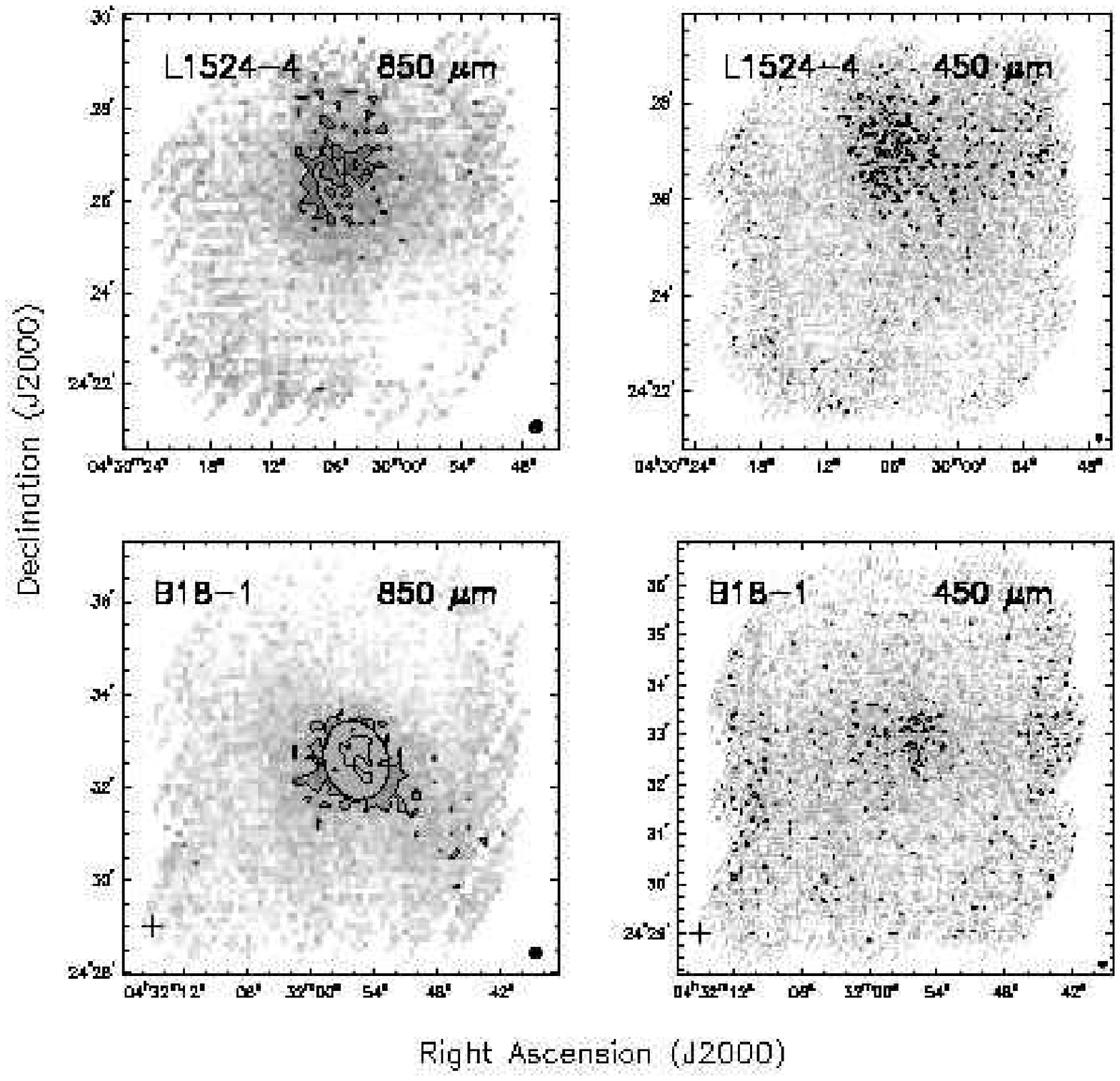} 
\figurenum{1}
\figcaption{continued...}
\end{figure}

\clearpage 

\begin{figure}
\plotone{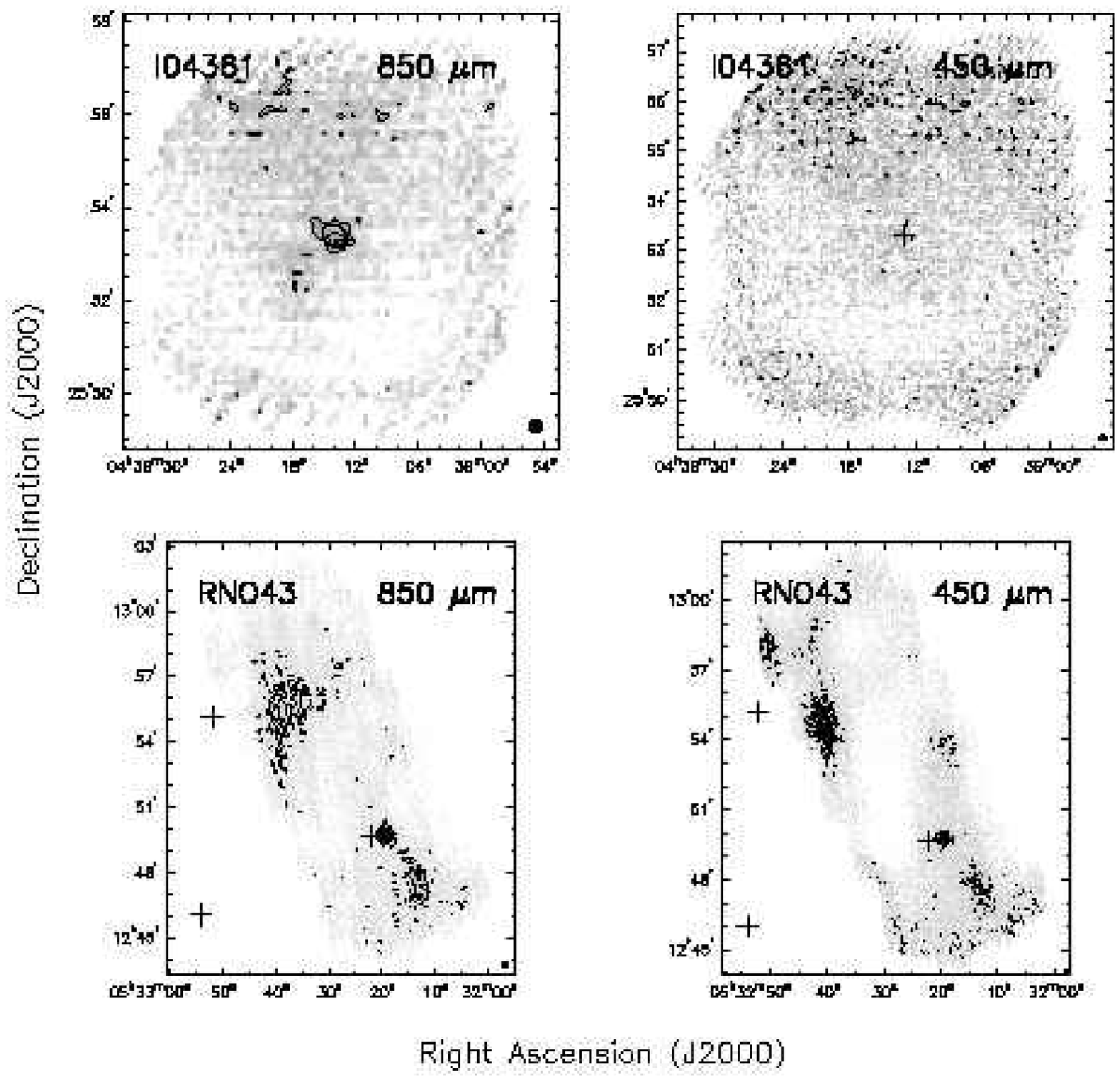} 
\figurenum{1}
\figcaption{continued...}
\end{figure}

\begin{figure}
\plotone{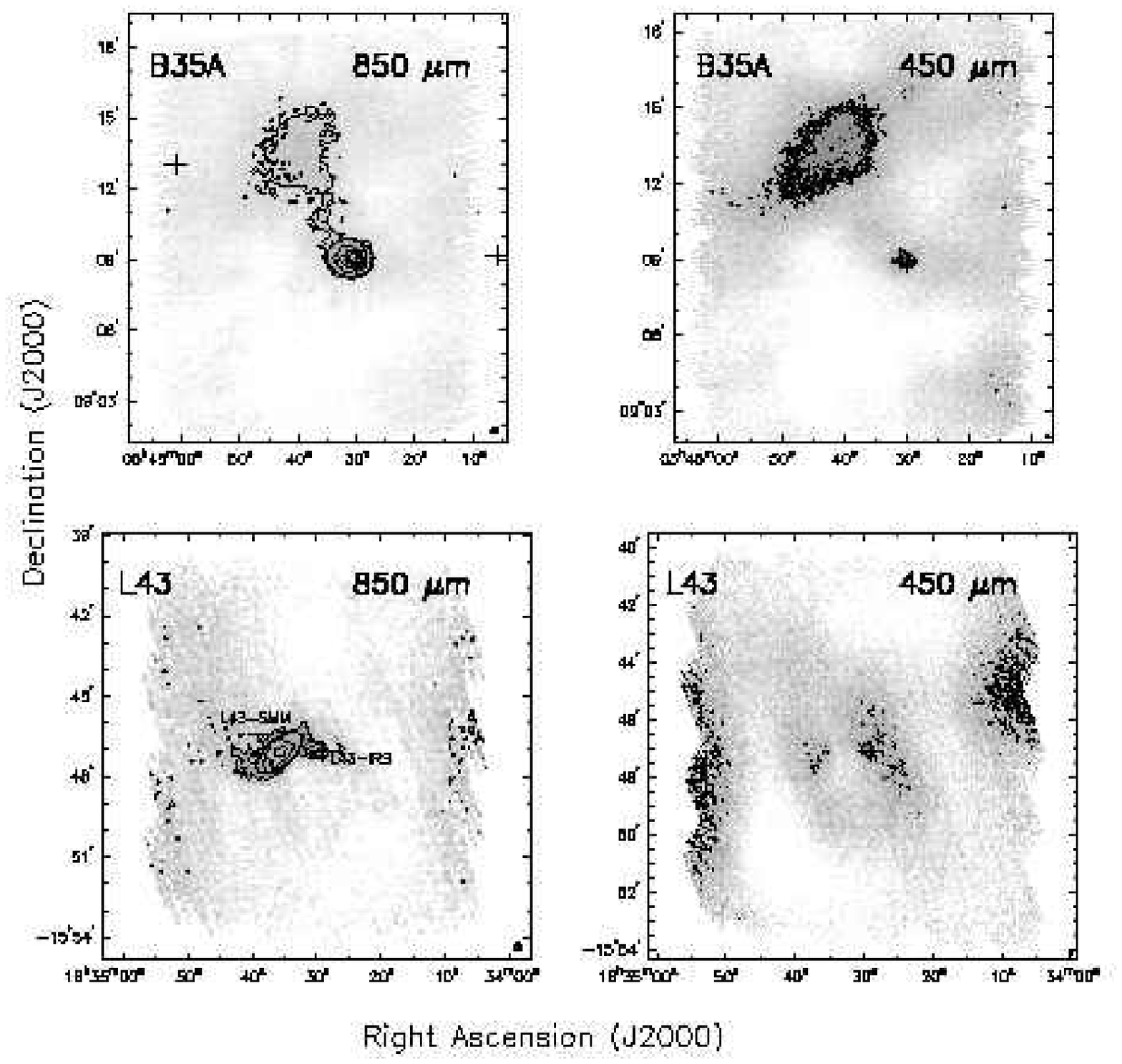} 
\figurenum{1}
\figcaption{continued...}
\end{figure}

\begin{figure}
\plotone{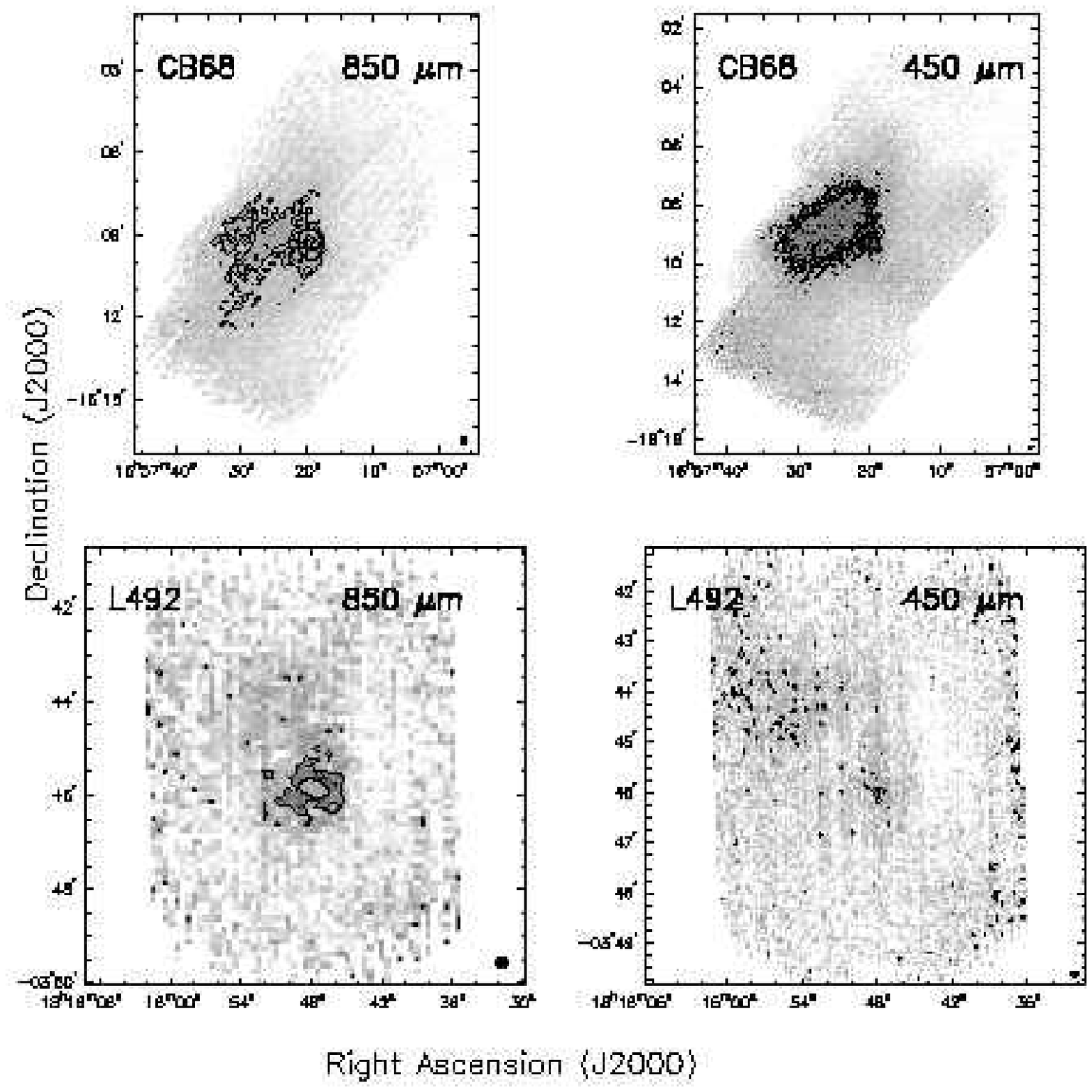} 
\figurenum{1}
\figcaption{continued...}
\end{figure}

\begin{figure}
\plotone{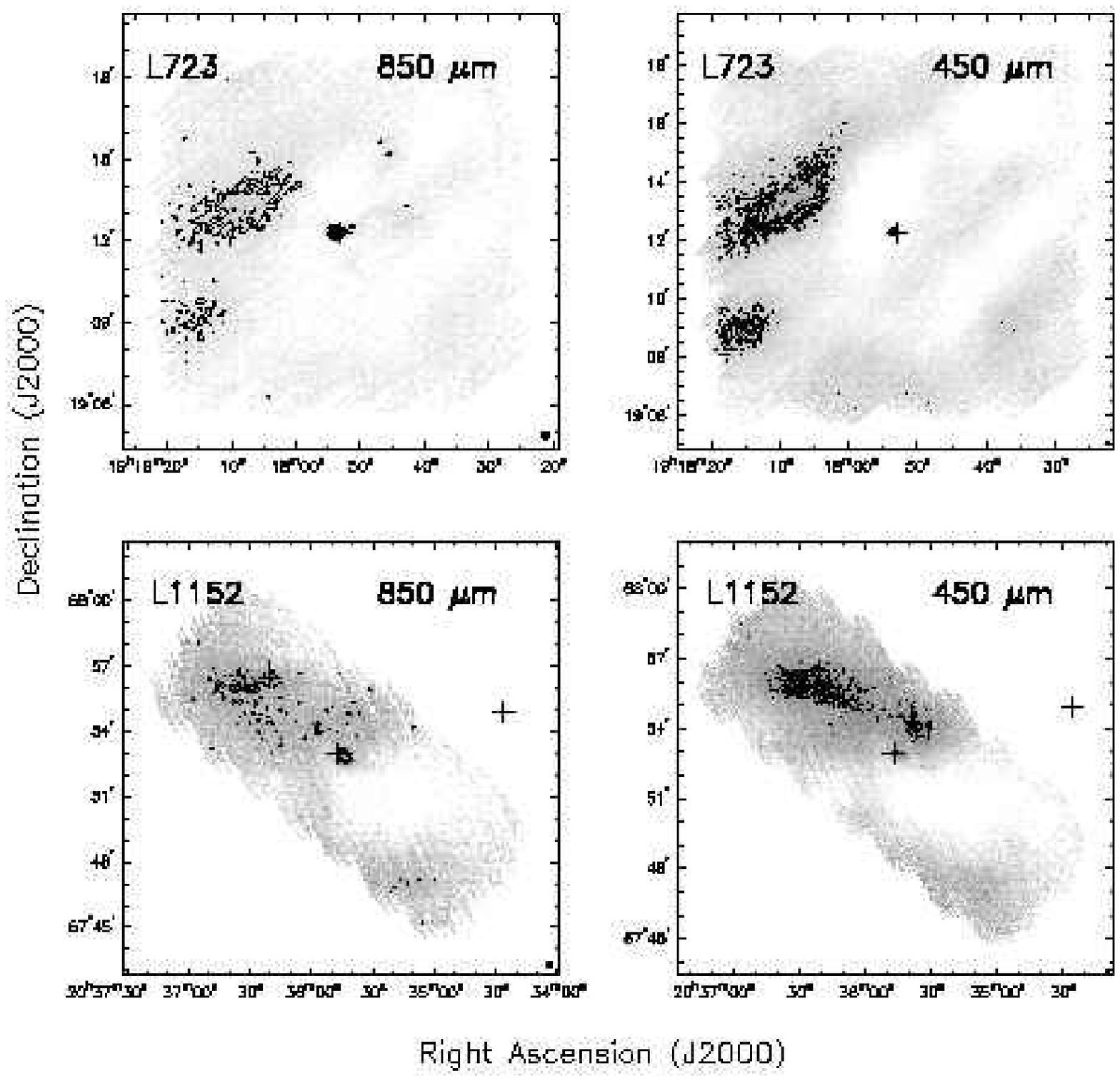} 
\figurenum{1}
\figcaption{continued...}
\end{figure}

\begin{figure}
\plotone{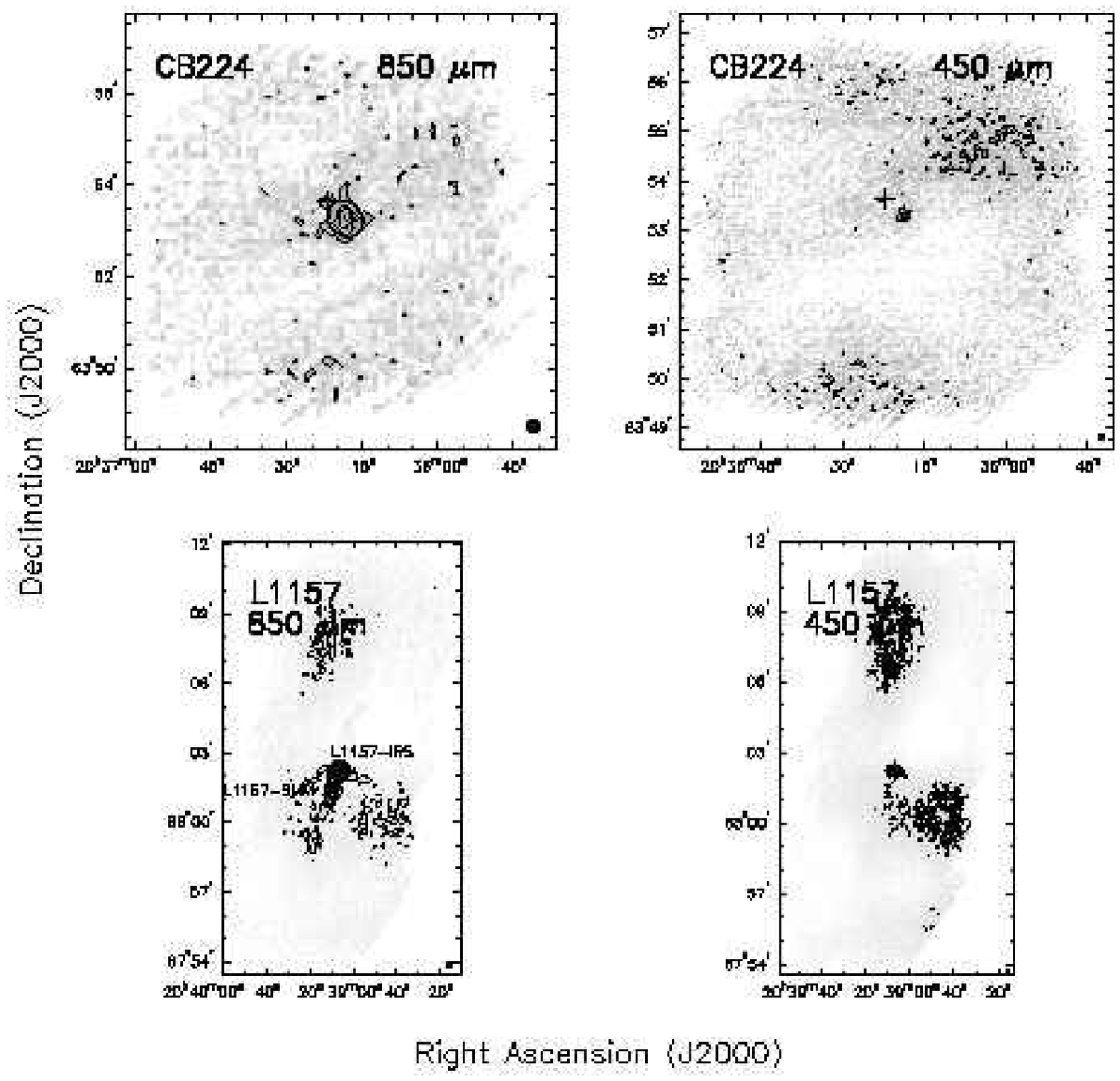}
\figurenum{1}
\figcaption{continued...}
\end{figure}

\begin{figure}
\plotone{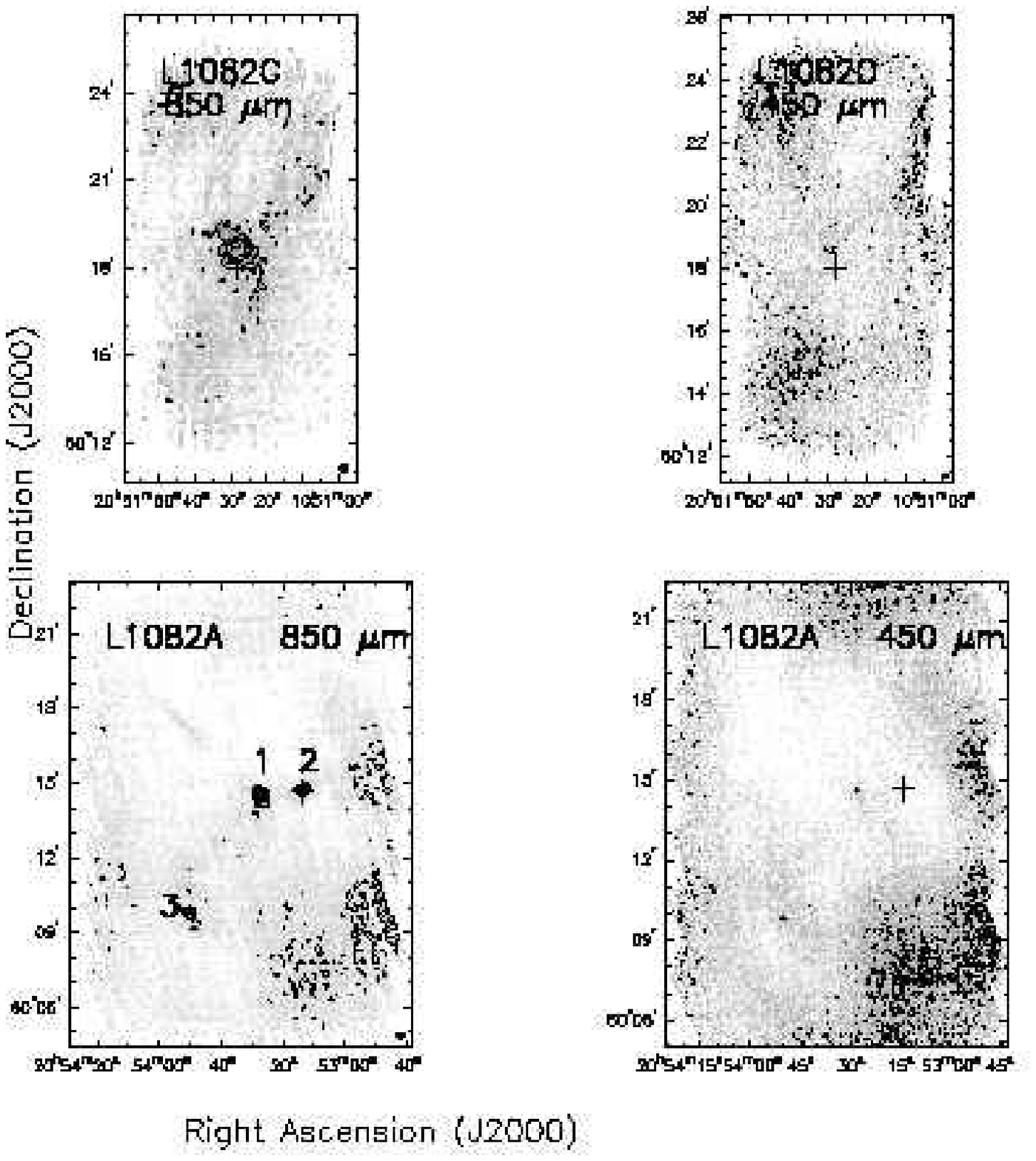}
\figurenum{1}
\figcaption{continued...}
\end{figure}

\begin{figure}
\plotone{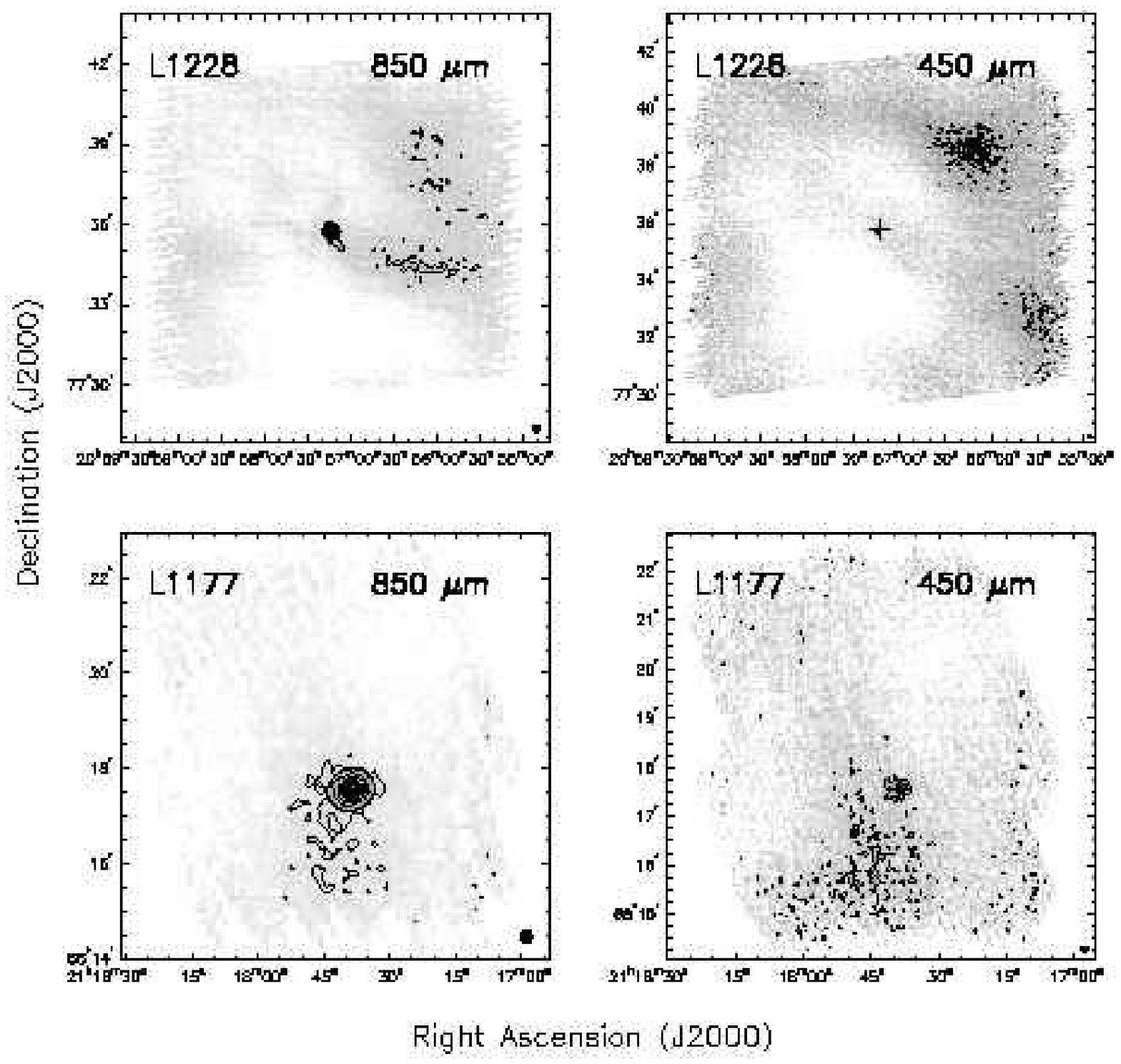}
\figurenum{1}
\figcaption{continued...}
\end{figure}

\begin{figure}
\plotone{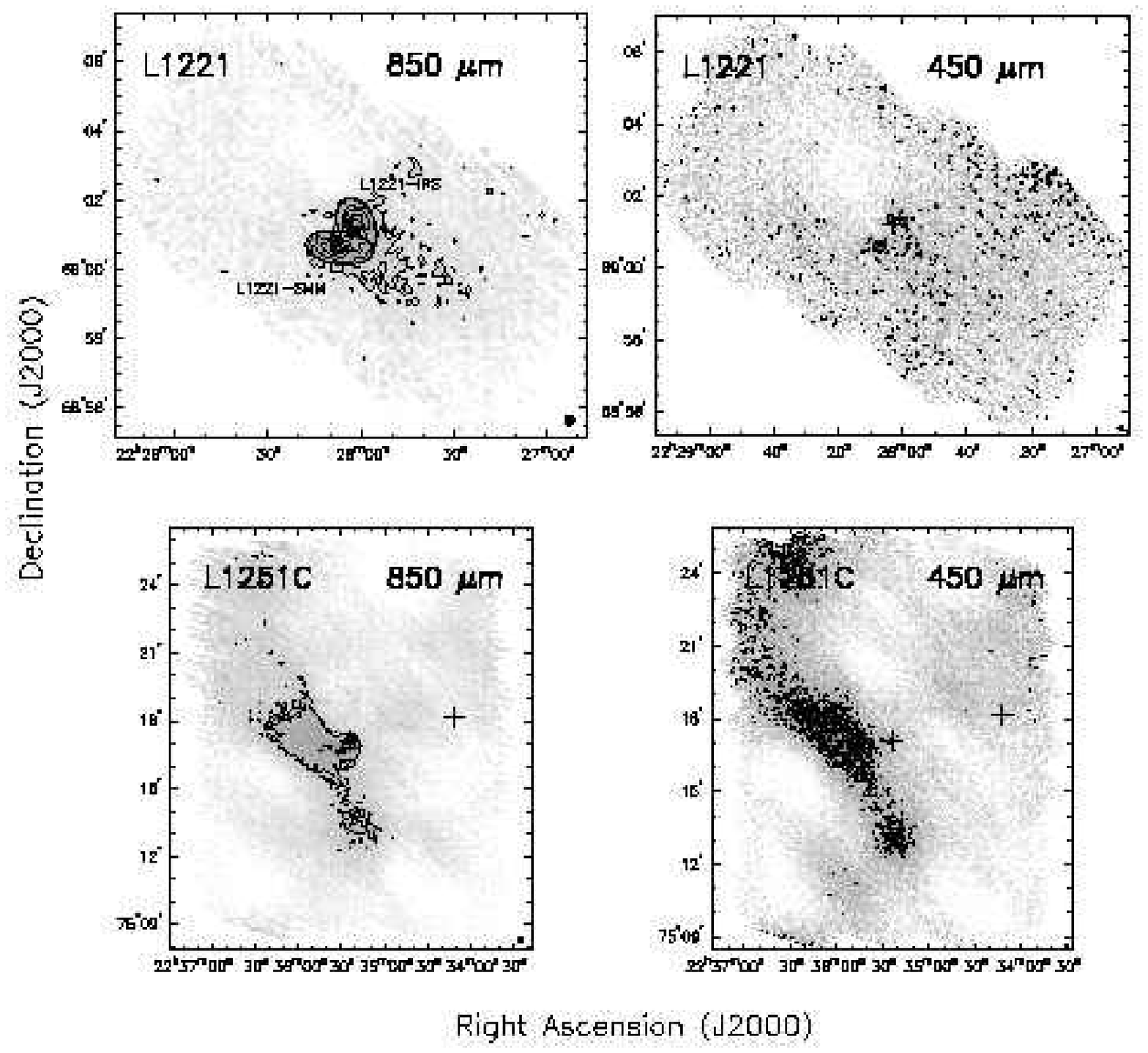}
\figurenum{1}
\figcaption{continued...}
\end{figure}

\begin{figure}
\plotone{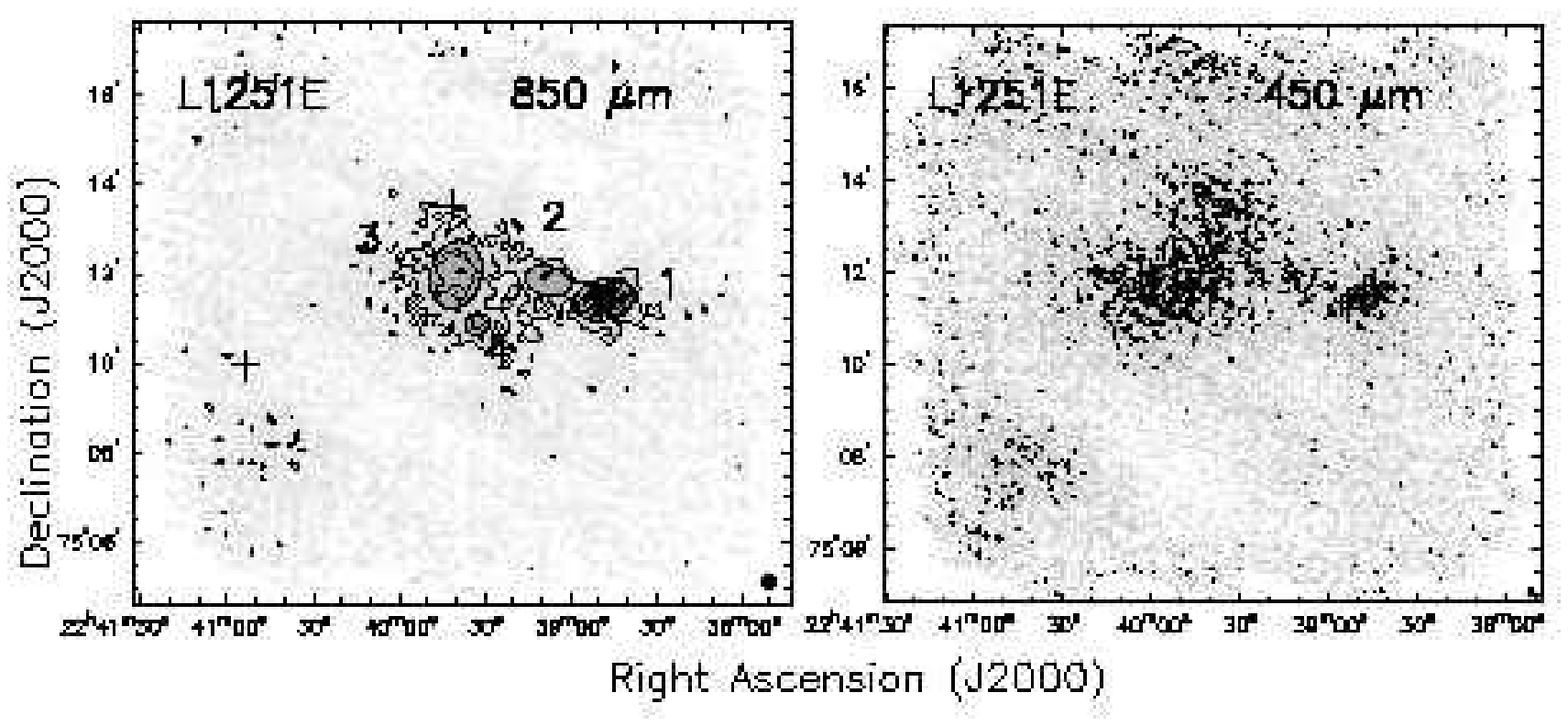}
\figurenum{1}
\figcaption{continued...}
\end{figure}

\clearpage

\clearpage

\clearpage

\clearpage
\begin{deluxetable}{llrlccccccccc}
\tabletypesize{\scriptsize}
\rotate
%\footnotesize
\tablecolumns{13}
\tablecaption{Sources\label{table-sources}}
\tablewidth{0pt} 
\tablehead{
\colhead{Source Name}                  &
\colhead{RA}                           &
\colhead{Dec}                          &
\colhead{Association}                  &
\colhead{Dist.}                     &
\colhead{Ref\tablenotemark{b}}         &
\colhead{IRAS\tablenotemark{a}}        &
\colhead{SCUBA}                        &
\colhead{SST\tablenotemark{c}}         &
\colhead{3-$\sigma$ (850$\mu$m)}       &
\colhead{3-$\sigma$ (450$\mu$m)}       &
\colhead{$\tau_{850}$}                 &
\colhead{$\tau_{450}$}                 \\
\colhead{}                             &
\colhead{(J2000)}                        &
\colhead{(J2000)}                        &
\colhead{}                             &
\colhead{(pc)}                           &
\colhead{}                             &
\colhead{(Y/N)}                        &
\colhead{(Y/N)}                        &
\colhead{(Y/N)}                        &
\colhead{(mJy/beam)}                           &
\colhead{(mJy/beam)}                           &
\colhead{}                             &
\colhead{}                             

}
\startdata 
    
\bf      PER4B & 03$^h$29$^m$17$\fs$9 & 31$\arcdeg$27$\arcmin$31$\arcsec$ &  Perseus    & 250 & 1 & NNY & YYY & P & 220 & 2500 & 0.4 & 2.4 \\
\bf      PER5 & 03$^h$29$^m$51$\fs$6 & 31$\arcdeg$39$\arcmin$04$\arcsec$ &  Perseus     & 250 & 1 &  Y  & Y & P & 190 & 3900 & 0.44 & 3.0 \\
\bf      PER6 & 03$^h$30$^m$14$\fs$9 & 30$\arcdeg$23$\arcmin$49$\arcsec$ &  Perseus     & 250 & 1 &  Y  & Y & P & 150 & 6500 & 0.44 & 3.0 \\
2MASS0347392 & 03$^h$47$^m$37$\fs$6 & 31$\arcdeg$19$\arcmin$27$\arcsec$ & Perseus               & 250 & 1 &  N  & Y & N & 110 & 4200 & 0.49 & 3.2 \\
\bf IRAM04191+1522 & 04$^h$21$^m$56$\fs$9 & 15$\arcdeg$29$\arcmin$47$\arcsec$ &  Taurus         & 140 & 2 &  NY & YY & Y & 120 & 800 & 0.26 & 1.3 \\
\bf    L1521F & 04$^h$28$^m$39$\fs$8 & 26$\arcdeg$51$\arcmin$35$\arcsec$ &  Taurus      & 140 & 2 &  N  & Y & Y & 190 & 3100 & 0.29 & 1.6 \\
\bf   L1524-4 & 04$^h$30$^m$05$\fs$7 & 24$\arcdeg$25$\arcmin$16$\arcsec$ &  Taurus      & 140 & 2 &  N  & Y & Y & 200 & 1700 & 0.39 & 1.7 \\
\bf     B18-1 & 04$^h$31$^m$57$\fs$7 & 24$\arcdeg$32$\arcmin$30$\arcsec$ &  Taurus      & 140 & 2 &  N  & Y & Y & 190 & 5900 & 0.44 & 2.7 \\
    IRAS 04361+2547 & 04$^h$39$^m$13$\fs$9 & 25$\arcdeg$53$\arcmin$21$\arcsec$ &  Taurus   & 140 & 2 &  Y  & Y & N & 180 & 8600 & 0.44 & 3.0 \\
     RNO43 & 05$^h$32$^m$27$\fs$9 & 12$\arcdeg$53$\arcmin$11$\arcsec$ &                         & 400 & 3 &  Y  & Y & N & 170 & 1800 & 0.22 & 1.2 \\
\bf      B35A & 05$^h$44$^m$37$\fs$0 & 09$\arcdeg$10$\arcmin$13$\arcsec$ &              & 400 & 3 &  Y  & Y & Y & 220 & 2000 & 0.19 & 0.9 \\
\bf       L43 & 16$^h$34$^m$30$\fs$2 & -15$\arcdeg$46$\arcmin$56$\arcsec$ & Ophiuchus   & 125 & 4 & YN & YY & Y & 240 & 2700 & 0.27 & 1.5 \\
\bf      CB68 & 16$^h$57$^m$20$\fs$5 & -16$\arcdeg$09$\arcmin$02$\arcsec$ & Ophiuchus   & 125 & 4 &  Y  & Y & Y & 370 & 4100 & 0.29 & 1.3 \\
\bf      L492 & 18$^h$15$^m$49$\fs$0 & -03$\arcdeg$45$\arcmin$30$\arcsec$ &             & 270 & 5 &  N  & Y & Y & 120 & 500 & 0.15 & 0.6 \\
\bf      L723 & 19$^h$17$^m$53$\fs$2 & 19$\arcdeg$12$\arcmin$17$\arcsec$ &              & 300 & 6 &  Y  & Y & Y & 190 & 1400 & 0.24 & 1.2 \\
\bf     L1152 & 20$^h$35$^m$46$\fs$0 & 67$\arcdeg$52$\arcmin$42$\arcsec$ &              & 325 & 7 &  Y  & Y & Y & 350 & 2700 & 0.2 & 0.9 \\
     CB224 & 20$^h$36$^m$20$\fs$3 & 63$\arcdeg$52$\arcmin$55$\arcsec$ &                         & 400 & 8 &  Y  & Y & N & 120 & 1000 & 0.19 & 1.0 \\
     L1157 & 20$^h$39$^m$06$\fs$2 & 68$\arcdeg$02$\arcmin$42$\arcsec$ &                         & 325 & 7 & YN & YY & N & 160 & 1100 & 0.16 & 0.7 \\
    L1082C & 20$^h$51$^m$27$\fs$6 & 60$\arcdeg$18$\arcmin$35$\arcsec$ &                         & 400 & 8 &  Y  & Y & N & 140 & 2500 & 0.34 & 2.1 \\
    L1082A & 20$^h$53$^m$34$\fs$0 & 60$\arcdeg$13$\arcmin$41$\arcsec$ &                 & 400 & 8 & YYN & YYY & N & 160 & 1400 & 0.22 & 1.1 \\
\bf     L1228 & 20$^h$57$^m$11$\fs$8 & 77$\arcdeg$35$\arcmin$48$\arcsec$ &              & 200 & 9 &  Y  & Y & Y & 350 & 5900 & 0.26 & 1.5 \\
     L1177 (CB230) & 21$^h$17$^m$43$\fs$0 & 68$\arcdeg$18$\arcmin$24$\arcsec$ &                 & 288 & 7 &  Y  & Y & N & 180 & 1600 & 0.26 & 1.2 \\
\bf     L1221 & 22$^h$28$^m$03$\fs$0 & 69$\arcdeg$01$\arcmin$12$\arcsec$ &              & 250 & 10 & YN & YY & Y & 160 & 2400 & 0.31 & 1.8 \\
\bf    L1251C & 22$^h$35$^m$24$\fs$0 & 75$\arcdeg$17$\arcmin$09$\arcsec$ &              & 300 & 11 &  Y  & Y & Y & 330 & 4800 & 0.26 & 1.5 \\
\bf    L1251E & 22$^h$38$^m$20$\fs$7 & 75$\arcdeg$11$\arcmin$03$\arcsec$ &              & 300 & 11 & YNYN & YYYY & Y & 260 & 5300 & 0.29 & 1.6 \\
\hline
\bf     PER7A & 03$^h$32$^m$26$\fs$6 & 30$\arcdeg$59$\arcmin$57$\arcsec$ &  Perseus     & 250 & 1 & Y  & N & P & 200 & 1800 & 0.37 & 2.2 \\
\bf     PER7B & 03$^h$33$^m$15$\fs$3 & 30$\arcdeg$59$\arcmin$54$\arcsec$ &  Perseus     & 250 & 1 & N & N & P & 140 & 1500 & 0.35 & 2.1 \\
\bf  L1521B-2 & 04$^h$23$^m$37$\fs$0 & 26$\arcdeg$40$\arcmin$06$\arcsec$ &  Taurus      & 140 & 2 & N & N & Y & 190 & 2300 & 0.4 & 2.4 \\
\bf   L1521-2 & 04$^h$29$^m$31$\fs$8 & 26$\arcdeg$59$\arcmin$59$\arcsec$ &  Taurus      & 140 & 2 & N & N & Y & 260 & 1600 & 0.39 & 1.7 \\
\bf     B18-5 & 04$^h$35$^m$53$\fs$0 & 24$\arcdeg$09$\arcmin$32$\arcsec$ &  Taurus      & 140 & 2 & N & N & Y & 180 & 9500 & 0.51 & 3.8 \\
\bf    TMC1-2 & 04$^h$41$^m$10$\fs$0 & 25$\arcdeg$49$\arcmin$28$\arcsec$ &  Taurus      & 140 & 2 & Y  & N & Y & 320 & 4800 & 0.55 & 2.7 \\
\bf    TMC1-1 & 04$^h$41$^m$44$\fs$0 & 25$\arcdeg$42$\arcmin$22$\arcsec$ &  Taurus      & 140 & 2 & N & N & Y & 190 & 24900 & 0.38 & 2.9 \\
      CB28 & 05$^h$06$^m$16$\fs$0 & -03$\arcdeg$56$\arcmin$29$\arcsec$ &                        & 450 & 12 & Y  & N & N & 350 & 6400 & 0.25 & 1.7 \\
DCld 253.6+02.9 & 08$^h$28$^m$44$\fs$0 & -33$\arcdeg$45$\arcmin$12$\arcsec$ &           & 450 & 13 & Y  & N & N & 330 & 7400 & 0.22 & 1.2 \\
\bf     L1772 & 17$^h$19$^m$32$\fs$7 & -26$\arcdeg$44$\arcmin$4$\arcsec$0 & Ophiuchus   & 125 & 4 & N & N & Y & 260 & 2900 & 0.26 & 1.5 \\
       L55 & 17$^h$22$^m$58$\fs$0 & -23$\arcdeg$57$\arcmin$18$\arcsec$ & Ophiuchus      & 125 & 4 & N & N & N & 280 & 3100 & 0.25 & 1.2 \\
\bf       B72 & 17$^h$23$^m$46$\fs$3 & -23$\arcdeg$41$\arcmin$20$\arcsec$ & Ophiuchus   & 125 & 4 & N & N & Y & 340 & 4100 & 0.32 & 1.9 \\
\bf    L1155C & 20$^h$43$^m$06$\fs$0 & 67$\arcdeg$49$\arcmin$56$\arcsec$ &              & 325 & 7 & Y  & N & Y & 210 & 1700 & 0.21 & 1.0 \\

\enddata \tablenotetext{a}{We search a 1$\arcmin$ radius for
detections by IRAS.  The IRAS sources must meet the criteria
presented by \citet{lee99}. More than one symbol implies that more
than one submillimeter core was found.}
            
\tablenotetext{b}{(1)\citet{enoch06},(2)\citet{elias78},(3)\citet{murden77},(4)\citet{degeus89},(5)\citet{straizys03},
(6)\citet{goldsmith84},(7)\citet{straizys92},(8)\citet{dobashi94},(9)\citet{kun98},(10)\citet{yonekura97},
(11)\citet{kun93},(12)\citet{ogura98},(13)\citet{woermann01}}

\tablenotetext{c}{A Y means that the core was observed by Spitzer (SST)
as part of the c2d cores
program, a N means that it was in the original cores program, but was cut;
and a P means that it was observed as part of the Perseus cloud map.
}
                                             
\end{deluxetable}

\begin{deluxetable}{lllcccccccccc}
\tabletypesize{\scriptsize}
%\rotate
\tablecolumns{13}
\tablecaption{Calibrators\label{table-calibrators}}
\tablewidth{0pt} 
\tablehead{
\colhead{Date}                          &
\colhead{Calibrator}                   &
\colhead{$\tau_{850}$}                   &
\colhead{$\tau_{450}$}                   &
\colhead{C$^{850}_{20}$\tablenotemark{a}}                 &
\colhead{C$^{850}_{40}$}                 &
\colhead{C$^{850}_{80}$}                 &
\colhead{C$^{850}_{120}$}                &
\colhead{C$^{450}_{20}$}                 &
\colhead{C$^{450}_{40}$}                 &
\colhead{C$^{450}_{80}$}                 &
\colhead{C$^{450}_{120}$}                &
\colhead{FWHM\tablenotemark{c} (\arcsec)}                
}
\startdata 
11 Jan 2002  & CRL618   & 0.26 & 1.31 & 79.0  & 41.4 & 31.5 & 24.6 & 38.3  & 22.2 & 10.5 & 6.0          &15.3   \\
11 Jan 2002  & Uranus   & 0.20 & 0.96 & 62.0  & 44.0 & 41.5 & 42.0 & 35.1  & 29.5 & 28.7 & 33.8         &15.6   \\
15 Jan 2002  & CRL618   & 0.40 & 2.37 & 61.8  & 41.5 & 31.9 & 25.0 & 23.6  & 12.6 & 5.6  & 3.1          &15.5   \\
15 Jan 2002  & Mars     & 0.35 & 2.14 & 62.8  & 47.3 & 43.6 & 42.9 & 34.2  & 25.9 & 21.5 & 19.8         &15.0   \\
15 Jan 2002  & Uranus   & 0.37 & 2.17 & 68.1  & 47.7 & 45.3 & 46.6 & 31.9  & 32.0 &119.2 & -21.5        &15.9   \\
17 Jan 2002  & CRL618   & 0.51 & 3.84 & 57.5  & 43.8 & 41.9 & 43.3 & 12.4  & 10.7 & 10.0 & 9.8          &15.5   \\
17 Jan 2002  & CRL618   & 0.44 & 3.03 & 76.9  & 47.8 & 43.6 & 42.5 & 24.4  & 15.9 & 7.8  & 5.5          &15.8   \\
17 Jan 2002  & Mars     & 0.52 & 3.24 & 80.6  & 49.6 & 45.1 & 43.9 & 29.5  & 21.8 & 17.1 & 15.3         &15.5   \\
18 Jan 2002  & CRL618   & 0.60 & 4.28 & 76.6  & 38.9 & 27.1 & 19.8 & 5.8   & 2.3  & 0.9  & 0.4          &16.6   \\
18 Jan 2002  & CRL618   & 0.60 & 4.28 & 67.0  & 48.0 & 47.9 & 55.5 & 16.6  & 17.7 & -8.4 & -2.6         &15.4   \\
18 Jan 2002  & Mars     & 0.55 & 2.69 & 72.0  & 49.3 & 43.1 & 41.0 & 66.3  & 46.5 & 36.0 & 31.4         &15.5   \\
\tableline
29 Oct 2002  & Jupiter\tablenotemark{b}  & 0.25 & 1.34 & 155.8  & 68.3 & 50.4 & 48.8 & 198.6  & 68.7 & 49.6 & 47.0      &\nodata\\ %not listed as a calibrator
29 Oct 2002  & Uranus   & 0.28 & 1.50 & 71.2  & 55.1 & 52.6 & 53.5 & 74.4  & 64.8 & 89.2 & 917.1        &15.7   \\
\tableline
23 Mar 2003  & IRC10216 \tablenotemark{b}& 0.18 & 0.97 & 54.6 &  30.9 & 25.4 & 26.2 & \nodata  & \nodata & \nodata & \nodata    &\nodata\\% not suggested to use as cal
23 Mar 2003  & IRC10216 & 0.17 & 0.88 &  52.6 & 29.4 & 24.0 & 24.6 &  \nodata  & \nodata & \nodata & \nodata    &\nodata\\% can't use for 450 cal
25 Mar 2003  & IRC10216 & 0.33 & 1.94 &  48.5 & 25.6 & 16.6 & 12.8 &  \nodata  & \nodata & \nodata & \nodata    &\nodata\\
26 Mar 2003  & IRC10216 & 0.35 & 2.11 &  50.9 & 28.8 & 26.2 & 27.8 & \nodata   & \nodata & \nodata & \nodata    &\nodata\\
26 Mar 2003  & Mars     & 0.25 & 1.45 &  61.7 & 42.3 & 39.0 & 38.7 & 54.6  & 40.9 & 36.4 & 36.5         &15.5   \\
27 Mar 2003  & IRC10216 & 0.30 & 1.57 &  54.0 & 29.8 & 23.6 & 22.5 &  \nodata  & \nodata & \nodata & \nodata    &\nodata\\
28 Mar 2003  & IRC10216 & 0.20 & 0.87 &  45.1 & 24.6 & 17.7 & 15.0 & \nodata   & \nodata & \nodata & \nodata    &\nodata\\ 
\tableline
20 May 2003  & Uranus   & 0.19 & 0.96 &  48.4 & 37.8 & 35.4 & 35.4 & 35.7  & 28.6 & 26.0 & 26.1         &14.4   \\
20 May 2003  & Uranus   & 0.16 & 0.72 &  56.2 & 39.8 & 37.8 & 37.5 & 50.2  & 38.7 & 33.8 & 34.3         &15.4   \\
22 May 2003  & Uranus   & 0.22 & 1.07 &  56.8 & 38.9 & 34.6 & 34.1 & 41.7  & 28.4 & 24.7 & 23.8         &16.0   \\
23 May 2003  & Uranus   & 0.24 & 1.22 &  52.4 & 40.0 & 37.3 & 36.8 & 51.0  & 41.6 & 37.1 & 35.6         &15.3   \\
27 May 2003  & Uranus   & 0.37 & 2.55 &  53.0 & 40.4 & 37.7 & 37.1 & 25.2  & 20.0 & 17.6 & 17.3         &15.1   \\
%28 May 2003  & Uranus   & 0.33 & 1.91 &  91.0 & 54.8 & 37.5 & 34.3 & 107.9 & 43.9 & 21.5 & 14.0        &17.8   \\
\tableline
\tableline
\multicolumn{10}{l}{Average Values (Standard Deviation)}        &       \\
             &           &   &  &  65(10) & 44(5) & 40(6)  & 39(9) &  36(18) & 28(15) & 28(31)  &\nodata        &$15\farcs5(0\farcs5)$\\

\enddata

\tablenotetext{a}{The calibration factor, in Jy/V, for a 40$\arcsec$
aperture. Also, we show the 80$\arcsec$ and 120$\arcsec$ calibration
factors.}  
\tablenotetext{b}{We do not include Jupiter or IRC10216 in the average
calibration factor.}
\tablenotetext{c}{We give the FWHM of the beam at 850 $\mu$m. We
deconvolve the measured beam size for the planet diameter; Uranus and
Mars have semi-diameters of 1\farcs72 and 3\farcs67, respectively}

\end{deluxetable}

%Table 3
\begin{deluxetable}{lccccccc}
%\tabletypesize{\scriptsize}
\tablecolumns{8}
\tablecaption{Source Properties at 3-$\sigma$ Contours\label{table-properties}}
\tablewidth{0pt} 
\tablehead{
\colhead{Source}                                &
\colhead{RA}                                    &
\colhead{Dec}                                   &
\colhead{Major}                                 &
\colhead{Minor}                                 &
\colhead{Aspect}                                &
\colhead{$S_\nu$ \tablenotemark{a}}             &
\colhead{$M_{core}$\tablenotemark{b}}           \\
\colhead{}                                      &
\colhead{J2000}                                 &
\colhead{J2000}                                 &
\colhead{Axis (\arcsec)}                        &
\colhead{Axis (\arcsec)}                        &
\colhead{Ratio}                                 &
\colhead{(Jy)}                                  &
\colhead{($M_\odot$)}                           
      
}
\startdata

\bf PER4-B   &  03$^h$29$^m$17$\fs$2  &  31$\arcdeg$27$\arcmin$53$\arcsec$               &  22  &  13  &  1.6  &  0.3(0.1) & 0.2(0.1)   \\
\bf PER4-C   &  03$^h$29$^m$18$\fs$2  &  31$\arcdeg$25$\arcmin$14 $\arcsec$              &  29  &  20  &  1.5  &  0.5(0.2) & 0.3(0.1)   \\
\bf      PER4-A   &  03$^h$29$^m$25$\fs$3  &  31$\arcdeg$28$\arcmin$20$\arcsec$          &  17  &  17  &  1    &  0.3(0.1) & 0.2(0.1)   \\
\bf       PER5   &  03$^h$29$^m$51$\fs$8  &  31$\arcdeg$39$\arcmin$11$\arcsec$           &  80  &  54  &  1.5  &  2.4(0.7) & 1.3(0.4)   \\
      2MASS0347392   &  03$^h$47$^m$39$\fs$4  &  31$\arcdeg$19$\arcmin$08$\arcsec$       &  27  &  18  &  1.5  &  0.2(0.1) & 0.1(0.03)   \\
 \bf IRAM 04191+1522   &  04$^h$21$^m$56$\fs$8  &  15$\arcdeg$29$\arcmin$51 $\arcsec$    &  75  &  46  &  1.6  &  1.4(0.4) & 0.2(0.1)   \\
 \bf IRAS 04191+1523   &  04$^h$22$^m$00$\fs$6  &  15$\arcdeg$30$\arcmin$26$\arcsec$     &  38  &  30  &  1.2  &  0.6(0.2) & 0.1(0.03)   \\
\bf     L1521F   &  04$^h$28$^m$39$\fs$7  &  26$\arcdeg$51$\arcmin$51$\arcsec$           &  142 &  87  &  1.6  &  5.9(1.8) & 1.0(0.3)   \\
\bf    L1524-4   &  04$^h$30$^m$03$\fs$0  &  24$\arcdeg$26$\arcmin$39$\arcsec$           &  9   &  9   &  1    &  0.1(0.03)& 0.02(0.01)   \\
\bf     B18-1   &  04$^h$31$^m$55$\fs$8  &  24$\arcdeg$32$\arcmin$35$\arcsec$            &  104 &  82  &  1.3  &  4.1(1.2) & 0.7(0.2)   \\
    IRAS 04361+2547   &  04$^h$39$^m$14$\fs$0  &  25$\arcdeg$53$\arcmin$23 $\arcsec$     &  34  &  27  &  1.2  &  0.7(0.2) & 0.1(0.04)   \\
     RNO43   &  05$^h$32$^m$19$\fs$4  &  12$\arcdeg$49$\arcmin$41$\arcsec$               &  43  &  38  &  1.1  &  1.8(0.5) & 2.6(0.8)   \\
\bf       B35A   &  05$^h$44$^m$31$\fs$1  &  09$\arcdeg$09$\arcmin$04$\arcsec$           &  119 &  95  &  1.3  &  8.3(2.5) & 12.0(3.6)   \\
\bf        L43-IRS   &  16$^h$34$^m$30$\fs$1  &  -15$\arcdeg$46$\arcmin$59 $\arcsec$     &  55  &  33  &  1.7  &  1.8(0.5) & 0.3(0.1)   \\
\bf        L43-SMM   &  16$^h$34$^m$35$\fs$8  &  -15$\arcdeg$47$\arcmin$07$\arcsec$      &  123 &  58  &  2.1  &  5.7(1.7) & 0.8(0.2)   \\
\bf       CB68   &  16$^h$57$^m$19$\fs$6  &  -16$\arcdeg$09$\arcmin$20$\arcsec$          &  68  &  51  &  1.3  &  3.9(1.2) & 0.5(0.2)   \\
\bf       L492   &  18$^h$15$^m$47$\fs$8  &  -03$\arcdeg$45$\arcmin$51$\arcsec$          &  39  &  22  &  1.8  &  0.4(0.1) & 0.3(0.1)   \\
 \bf      L723   &  19$^h$17$^m$53$\fs$7  &  19$\arcdeg$12$\arcmin$19$\arcsec$           &  32  &  27  &  1.2  &  1.5(0.5) & 1.2(0.4)   \\
 \bf     L1152   &  20$^h$35$^m$44$\fs$3  &  67$\arcdeg$52$\arcmin$52$\arcsec$           &  38  &  21  &  1.8  &  1(0.3)   & 1.0(0.3)   \\
     CB224   &  20$^h$36$^m$18$\fs$5  &  63$\arcdeg$53$\arcmin$16$\arcsec$               &  51  &  42  &  1.2  &  1(0.3)   & 1.4(0.4)   \\
     L1157-IRS   &  20$^h$39$^m$06$\fs$3  &  68$\arcdeg$02$\arcmin$14$\arcsec$           &  52  &  48  &  1.1  &  3(0.9)   & 2.9(0.9)   \\
     L1157-SMM   &  20$^h$39$^m$09$\fs$7  &  68$\arcdeg$01$\arcmin$26$\arcsec$           &  95  &  33  &  2.9  &  2.1(0.6) & 2.0(0.6)   \\
    L1082C   &  20$^h$51$^m$28$\fs$2  &  60$\arcdeg$18$\arcmin$38$\arcsec$             &  50  &  45  &  1.1  &  1.3(0.4) & 1.9(0.6)   \\
    L1082A-2   &  20$^h$53$^m$13$\fs$7  &  60$\arcdeg$14$\arcmin$42$\arcsec$             &  26  &  22  &  1.2  &  0.5(0.2) & 0.7(0.3)   \\
    L1082A-1   &  20$^h$53$^m$27$\fs$6  &  60$\arcdeg$14$\arcmin$34$\arcsec$             &  38  &  28  &  1.3  &  1.1(0.3) & 1.6(0.6)   \\
    L1082A-3   &  20$^h$53$^m$50$\fs$2  &  60$\arcdeg$09$\arcmin$47$\arcsec$             &  27  &  21  &  1.3  &  0.5(0.2) & 0.7(0.3)   \\
\bf      L1228   &  20$^h$57$^m$13$\fs$4  &  77$\arcdeg$35$\arcmin$43$\arcsec$           &  34  &  29  &  1.2  &  2(0.6)   & 0.7(0.2)   \\
     L1177   &  21$^h$17$^m$38$\fs$8  &  68$\arcdeg$17$\arcmin$33$\arcsec$              &  59  &  52  &  1.1  &  3.1(0.9) & 2.3(0.7)   \\
L1251C\tablenotemark{c} & 22$^h$35$^m$24$\fs$6 & 75$\arcdeg$17$\arcmin$04$\arcsec$       & 35   & 31   & 1.1   & 2.2(0.7)  & 1.8(0.5)   \\
\bf      L1221-IRS   &  22$^h$28$^m$01$\fs$4  &  69$\arcdeg$01$\arcmin$18$\arcsec$       &  95  &  63  &  1.5  &  4.7(1.4) & 2.6(0.8)   \\
\bf     L1221-SMM   &  22$^h$28$^m$06$\fs$9  &  69$\arcdeg$00$\arcmin$38$\arcsec$        &  117 &  57  &  2.1  &  4.7(1.4) & 2.6(0.8)   \\
\bf    L1251E-1   &  22$^h$38$^m$48$\fs$1  &  75$\arcdeg$11$\arcmin$26$\arcsec$          &  83  &  55  &  1.5  &  6.4(1.9) & 5.2(1.6)   \\
\bf     L1251E-2   &  22$^h$39$^m$07$\fs$7  &  75$\arcdeg$11$\arcmin$51$\arcsec$         &  62  &  39  &  1.6  &  1.9(0.6) & 1.5(0.5)   \\
\bf     L1251E-4   &  22$^h$39$^m$32$\fs$3  &  75$\arcdeg$10$\arcmin$53$\arcsec$         &  33  &  28  &  1.2  &  0.9(0.3) & 0.7(0.2)   \\
\bf     L1251E-3   &  22$^h$39$^m$40$\fs$0  &  75$\arcdeg$11$\arcmin$59$\arcsec$         &  93  &  68  &  1.4  &  5.2(1.6) & 4.2(1.3)   \\

\enddata

\tablenotetext{a}{We give the 850 $\mu$m flux calculated for the
elliptical aperture described in this table.}

\tablenotetext{b}{The mass of the core is derived from the 850 $\mu$m
flux assuming an isothermal temperature of 15 K.}

\tablenotetext{c}{As discussed in the text, the properties of L1251C
were calculated with the analysis threshold set to 5-$\sigma$.}

\end{deluxetable}

\begin{deluxetable}{lcccccc}
\tabletypesize{\scriptsize}
\tablecolumns{7}
\tablecaption{Source Fluxes\label{table-fluxes}}
\tablewidth{0pt} 
\tablehead{
\colhead{Source Name}                        &
\colhead{$S_{20}^{850}$\tablenotemark{a}}                 &
\colhead{$S_{40}^{850}$}                 &
\colhead{$S_{80}^{850}$}                 &
\colhead{$S_{120}^{850}$}                 &
\colhead{$S_{20}^{450}$}                 &
\colhead{$S_{40}^{450}$}                 
}
\startdata 
%%
%%
%%  I have entered this data on 18 May 2005.  I've updated the aperture positions and
%%  have removed the 05/28 Uranus obs, which were bad.
%%
%%  I have entered this data on 26 May;  I removed 450 fluxes for those sources with
%%  detections weaker than 2-sigma
%%
\bf PER4-B    &   0.5(0.09)   &   0.7(0.1)   &   0.9(0.2)   &   0.6(0.3)   &   2.0(1.1)   &   2.8(1.6)   \\
\bf PER4-C    &   0.4(0.08)   &   0.9(0.1)   &   1.4(0.3)   &   2.0(0.5)   &   2.1(1.1)   &   4.4(2.4)   \\
\bf PER4-A    &   0.3(0.05)   &   0.6(0.1)   &   0.9(0.2)   &   0.9(0.3)   &   3.3(1.7)   &   7.7(4.2)   \\
\bf PER5    &   0.8(0.14)   &   1.3(0.2)   &   3.0(0.5)   &   4.9(1.2)   &   5.6(2.9)   &   11.9(6.4)   \\
2MASS0347392    &   0.2(0.04)   &   0.5(0.1)   &   0.9(0.2)   &   1.1(0.3)   &   \nodata            &   \nodata            \\
\bf IRAM 04191+1522    &   0.5(0.09)   &   0.9(0.1)   &   1.8(0.3)   &   2.8(0.7)   &   1.6(0.8)   &   1.1(0.7)   \\
\bf IRAS 04191+1523    &   0.4(0.07)   &   0.6(0.1)   &   1.2(0.2)   &   1.9(0.5)   &   1.4(0.7)   &   1.4(0.8)   \\
\bf L1521F    &   0.6(0.10)   &   1.5(0.2)   &   4.2(0.7)   &   6.8(1.6)   &   4.6(2.4)   &   13.3(7.2)   \\
\bf L1524-4    &   0.3(0.05)   &   0.6(0.1)   &   1.9(0.3)   &   3.8(0.9)   &   2.0(1.0)   &   5.6(3.0)   \\
\bf B18-1    &   0.3(0.04)   &   1.1(0.1)   &   3.4(0.6)   &   6.1(1.5)   &   9.1(4.6)   &   22.8(12.3)   \\
IRAS 04361+2547    &   0.5(0.08)   &   0.8(0.1)   &   1.5(0.3)   &   1.9(0.5)   &   \nodata\tablenotemark{b} &   \nodata            \\
RNO43    &   1.2(0.2)   &   1.7(0.3)   &   2.3(0.5)   &   2.7(0.7)   &   8.8(4.4)   &   13.8(7.5)   \\
\bf B35A    &   0.8(0.13)   &   2.6(0.3)   &   6.7(1.1)   &   10.0(2.4)   &   3.5(1.7)   &   12.2(6.6)   \\
\bf L43-IRS    &   0.6(0.10)   &   1.4(0.2)   &   2.9(0.5)   &   4.8(1.2)   &   5.9(3.0)   &   15.0(8.1)   \\
\bf L43-SMM    &   0.7(0.11)   &   2.0(0.2)   &   5.2(0.8)   &   8.2(2.0)   &   \nodata            &   \nodata            \\
\bf CB68    &   1.1(0.17)   &   2.3(0.3)   &   5.5(0.9)   &   9.4(2.3)   &   10.8(5.4)   &   27.8(14.9)   \\
\bf L492    &   0.2(0.03)   &   0.5(0.1)   &   1.4(0.2)   &   2.1(0.5)   &   0.7(0.3)   &   1.7(0.9)   \\
\bf L723    &   1.4(0.26)   &   1.4(0.3)   &   1.2(0.4)   &   0.3(0.4)   &   4.0(2.1)   &   0.7(0.9)   \\
\bf L1152    &   0.5(0.08)   &   1.2(0.2)   &   2.2(0.4)   &   2.9(0.7)   &   2.9(1.4)   &   8.3(4.4)   \\
CB224    &   0.4(0.07)   &   0.8(0.1)   &   1.5(0.3)   &   1.9(0.5)   &   2.3(1.2)   &   2.8(1.6)   \\
L1157-SMM    &   0.7(0.11)   &   1.2(0.2)   &   2.5(0.4)   &   5.7(1.4)   &   \nodata            &   \nodata            \\
L1157-IRS    &   2.2(0.39)   &   2.5(0.5)   &   3.7(0.8)   &   5.0(1.3)   &   9.7(4.9)   &   13.2(7.2)   \\
L1082C    &   0.3(0.05)   &   1.0(0.1)   &   2.2(0.4)   &   3.4(0.8)   &   1.9(1.0)   &   4.8(2.6)   \\
L1082A-2    &   0.4(0.07)   &   0.6(0.1)   &   0.8(0.2)   &   0.5(0.2)   &   \nodata            &   \nodata            \\
L1082A-1    &   0.8(0.15)   &   1.1(0.2)   &   1.5(0.3)   &   1.7(0.5)   &   1.1(0.6)   &   0.02(0.4)   \\
L1082A-3    &   0.5(0.08)   &   0.6(0.1)   &   1.1(0.2)   &   1.7(0.4)   &   1.7(0.9)   &   1.8(1.0)   \\
\bf L1228    &   1.2(0.20)   &   2.0(0.3)   &   3.5(0.7)   &   4.5(1.2)   &   3.6(2.1)   &   1.1(1.4)   \\
L1177     &   1.6(0.29)   &   2.4(0.4)   &   3.9(0.7)   &   5.3(1.3)   &   6.3(3.2)   &   10.7(5.8)   \\
L1251C & 0.9(0.2) & 2.3(0.3) & 5.2(0.9) & 8.7(2.1) & 8.9(4.5) & 16.4(8.9) \\
\bf L1221-IRS   &   1.0(0.16)   &   2.0(0.2)   &   4.9(0.8)   &   8.1(2.0)   &   4.9(2.5)   &   10.4(5.6)   \\
\bf L1221-SMM    &   1.2(0.19)   &   2.0(0.3)   &   4.5(0.8)   &   7.9(1.9)   &   6.5(3.3)   &   11.3(6.1)   \\
\bf L1251E-1    &   2.0(0.33)   &   4.0(0.5)   &   7.0(1.3)   &   9.5(2.3)   &   24.1(12.1)   &   49.5(26.6)   \\
\bf L1251E-2    &   0.5(0.08)   &   1.4(0.2)   &   3.5(0.6)   &   5.9(1.4)   &   \nodata            &   \nodata            \\
\bf L1251E-4    &   0.6(0.10)   &   1.2(0.2)   &   3.3(0.5)   &   6.6(1.6)   &   \nodata            &   \nodata            \\
\bf L1251E-3    &   0.5(0.09)   &   1.9(0.2)   &   5.2(0.8)   &   9.4(2.2)   &   \nodata            &   \nodata            \\

\enddata
\tablenotetext{a}{The flux, in Jy, for a 20$\arcsec$ aperture. The
remaining columns are fluxes in 40$\arcsec$, 80$\arcsec$, and
120$\arcsec$ apertures. We also give the 450 $\mu$m fluxes as measured
in 20$\arcsec$ and 40$\arcsec$ apertures.}

\tablenotetext{b}{We require, at least, a 2-$\sigma$ detection to
report the flux measurement; several cores did not satisfy this
requirement in the 450 $\mu$m data.}

\end{deluxetable}

\clearpage

\end{document}